\documentclass{article}
\pdfoutput=1
\usepackage{graphicx}
\usepackage[small,it]{caption}
\usepackage[margin=1in]{geometry}
\usepackage{amsmath, amsfonts}
\usepackage[style=numeric-comp,backend=bibtex,eprint=false,doi=false,isbn=false,url=false,sorting=none, maxbibnames=10,giveninits=true]{biblatex}
\renewbibmacro{in:}{%
  \ifentrytype{article}
    {}
    {\printtext{\bibstring{in}\intitlepunct}}}
\usepackage{epstopdf}
\usepackage{authblk}

\newcommand\blfootnote[1]{%
  \begingroup
  \renewcommand\thefootnote{}\footnote{#1}%
  \addtocounter{footnote}{-1}%
  \endgroup
}

\usepackage{multicol}
\usepackage{nonfloat}
\usepackage{color}
\usepackage{bm}
\usepackage{algorithm}
\usepackage{algpseudocode}
\usepackage{multirow}

\usepackage{tikz}
\usepackage{amsmath}
\usepackage{amssymb}
\usepackage{breqn}
\usepackage{etoolbox}
\usepackage{tabularx}
\usepackage{xparse}

\newrobustcmd*{\mysquare}[2]{\tikz{\draw[draw=#1,fill=#2] (0cm,0cm)
rectangle (0.2cm,0.2cm)}}
\newrobustcmd*{\mythicksquare}[2]{\tikz{\draw[line width=0.3mm,draw=#1,fill=#2] (0cm,0cm)
rectangle (0.2cm,0.2cm)}}
\newrobustcmd*{\mybarredsquare}[2]{\tikz{\draw[draw=#1,fill=#2] (0,0)
rectangle (0.2cm,0.2cm); \draw[draw=#1] (-0.1cm, 0.1cm) -- (0.3cm, 0.1cm)}}
\newrobustcmd*{\mythickbarredsquare}[2]{\tikz{\draw[line width=0.3mm,draw=#1,fill=#2] (0,0)
rectangle (0.2cm,0.2cm); \draw[draw=#1] (-0.1cm, 0.1cm) -- (0.3cm, 0.1cm)}}

\newrobustcmd*{\mycircle}[2]{\tikz{\draw[draw=#1,fill=#2] (0,0)
circle (0.1cm);}}
\newrobustcmd*{\mythickcircle}[2]{\tikz{\draw[line width=0.3mm,draw=#1,fill=#2] (0,0)
circle (0.1cm);}}
\newrobustcmd*{\mybarredcircle}[2]{\tikz{\draw[draw=#1,fill=#2] (0,0)
circle (0.1cm); \draw[draw=#1] (-0.2cm, 0.0cm) -- (0.2cm, 0.0cm)}}
\newrobustcmd*{\mythickbarredcircle}[2]{\tikz{\draw[line width=0.3mm,draw=#1,fill=#2] (0,0)
circle (0.1cm); \draw[draw=#1] (-0.2cm, 0.0cm) -- (0.2cm, 0.0cm)}}

\newrobustcmd*{\mytriangle}[2]{\tikz{\filldraw[draw=#1,fill=#2] (0.0cm,0.0cm) --
(0.2cm,0cm) -- (0.1cm,0.2cm) -- (0cm,0cm);}}
\newrobustcmd*{\mythicktriangle}[2]{\tikz{\filldraw[line width=0.3mm,draw=#1,fill=#2] (0.0cm,0cm) --
(0.2cm,0cm) -- (0.1cm,0.2cm) -- (0.0cm,0cm);}}
\newrobustcmd*{\mybarredtriangle}[2]{\tikz{\draw[draw=#1,fill=#2] (0,0) --
(0.2cm,0) -- (0.1cm,0.2cm) -- (0cm,0cm); \draw[draw=#1] (-0.1cm, 0.07cm) -- (0.3cm, 0.07cm)}}
\newrobustcmd*{\mythickbarredtriangle}[2]{\tikz{\draw[line width=0.3mm,draw=#1,fill=#2] (0,0) --
(0.2cm,0) -- (0.1cm,0.2cm) -- (0cm,0cm); \draw[draw=#1] (-0.1cm, 0.07cm) -- (0.3cm, 0.07cm)}}

\newrobustcmd*{\mythickline}[1]{\tikz{\draw[line width=0.3mm,draw=#1] (-0.15cm, 0.1cm) -- (0.15cm, 0.1cm);\draw[line width=0.3mm,draw=#1] (-0.0cm, 0.0cm);}}
\newrobustcmd*{\mythickdashedline}[1]{\tikz{\draw[line width=0.3mm,draw=#1] (-0.2, 0.1cm) -- (-0.1cm, 0.1cm); \draw[line width=0.3mm,draw=#1] (-0.0cm, 0.1cm) -- (0.1cm, 0.1cm); \draw[line width=0.3mm,draw=#1] (-0.0cm, 0.0cm);}}

\title{A Neural Network Inspired Formulation of Chemical Kinetics}
\date{}
\author{Shivam Barwey\thanks{Corresponding author. Email: \texttt{sbarwey@umich.edu}}}
\author{Venkat Raman}
\affil{\small{Department of Aerospace Engineering, University of Michigan, Ann Arbor, MI 48109, USA}}

\begin{document}

\maketitle

\begin{abstract}
\noindent
A method which casts the chemical source term computation into an artificial neural network (ANN)-inspired form is presented. This approach is well-suited for use on emerging supercomputing platforms that rely on graphical processing units (GPUs). The resulting equations allow for a GPU-friendly matrix-multiplication based source term estimation where the leading dimension (batch size) can be interpreted as the number of chemically reacting cells in the domain; as such, the approach can be readily adapted in high-fidelity solvers for which an MPI rank offloads the source term computation task for a given number of cells to the GPU. Though the exact ANN-inspired recasting shown here is optimal for GPU environments as-is, this interpretation allows the user to replace portions of the exact routine with trained, so-called approximate ANNs, where the goal of these approximate ANNs is to increase computational efficiency over the exact routine counterparts. Note that the main objective of this paper is not to use machine learning for developing models, but rather to represent chemical kinetics using the ANN framework. The end result is that little-to-no training is needed, and the GPU-friendly structure of the ANN formulation during the source term computation is preserved. The method is demonstrated using chemical mechanisms of varying complexity on both 0-D auto-ignition and 1-D channel detonation problems, and the details of performance on GPUs are explored.
\end{abstract}
\blfootnote{Preprint submitted to \textit{Combustion and Flame}.}

\section{Introduction}
Computational modeling is an integral component of the design of modern combustion devices. While there has been considerable growth in the physical understanding and the development of reliable yet computationally efficient models \cite{mindthegap,emerging_trends}, representation of complex chemical kinetics remains both a computational and modeling challenge. In particular, the use of detailed mechanisms that involve a hundred or more species and an even larger number of reactions in a turbulent flow configuration still remains out of reach \cite{jackie_proc_review}. While progress towards their use in canonical flows has been reported \cite{benedicte_sandia}, their use in simulation of complex geometries is still limited. When modeling turbulent combustion, manifold methods have overcome this computational issue by representing multi-step kinetics using a reduced-set of tracking variables such as mixture fraction and progress variable \cite{mueller_turnkey}. However, other combustion models such as the transported probability density function (PDF) approach \cite{popebook, ramanpitsch-sandia} or the linear-eddy model \cite{menon_lem} require detailed chemistry to be directly evolved. In this regard, methods and algorithms that allow detailed chemical processes to be included in such approaches are a critical requirement.

In the past, several approaches have been used to accelerate chemical source term computations. These include tabulation methods such as in-situ adaptive tabulation (ISAT) \cite{pope_isat} and the PRISM \cite{prism} approach. In these methods, the computationally expensive numerical integration of chemical source terms, which can be cast as a set of ordinary differential equations (ODEs), is replaced by a look-up table. In particular, ISAT builds a trust region in thermochemical composition space using a set of ellipsoids determined by the Jacobian of the source term. However, the cost of building and accessing such tables can become expensive, especially on modern high performance computers that are memory-limited and use extensive concurrency of computations to reach high throughput efficiency. An alternative approach, which is also the focus here, is based on artificial neural networks (ANNs) \cite{christo_ann, menon_ann,kempf_ann,kundu_ann,shivam_ftc}. Before discussing the specifics of the ANN for kinetics, it is necessary to describe a parallel trend in computing hardware.

The use of ANNs has driven the overall revolution of data sciences. A critical enabling tool for ANNs has been the development of hardware for machine learning. Due to the large application scope of artificial intelligence and machine learning, modern high-performance computing (HPC) revolves around the usage of graphics processing units (GPUs) or similar accelerators whose architectures enable fast algorithm execution in single-instruction, multiple-thread (SIMT) environments \cite{nickolls2010gpu}. Additionally, the compute power of modern day HPCs is increasingly being dominated by GPUs due to their power efficiency and high theoretical peak performance. It is has become crucial for the CFD community to adapt to these changes, though a central issue revolves around the re-interpretation and re-design of traditional algorithms that have been around for decades into a GPU-optimal scope \cite{niemeyer2014recent}. In general, GPUs operate differently from CPUs, requiring algorithmic implementations to be vastly altered in order to leverage their specific hardware architecture. To this end, approaches for GPU-offloading of kinetics have been explored in detail in recent years to good success \cite{niemeyer2014,niemeyer2018,rigopoulos_gpu}, and their implementation into high-fidelity parallel solvers has also been demonstrated \cite{sankaran_solver}. These approaches traditionally rely on translation of the exact equations for kinetics and time-integration methods into the GPU environment.

However, given that ANN libraries that take advantage of specific GPU capabilities already exist, ensuring a readily interpretable and accurate ANN representation of chemistry will allow kinetics calculations to be performed efficiently. The focus of this work is to develop such an ANN-based framework for GPUs. To provide context, prior use of ANNs for chemical kinetics computations can be placed into two categories, each providing a different ANN representation and levels of interpretability. The first is to replace both the source term computation and the time integration step with a single trained ANN that takes thermochemical state as input and outputs the same state at a future time step (ANN as a time integrator) \cite{sharma2020deep,blasco_ann,blasco_ann2}. Assuming a relatively simple architecture, this approach is attractive due to its speed: if it works, both source term recovery and time integration is captured in a single efficient pass through the ANN. However, the downside is that this approach relies heavily on sampling a high-dimensional data space during the training process, which is either prohibitive for high-dimensional mechanisms or requires involved sampling procedures that rely on slow manifold theory. Furthermore, because the time integration is treated within the ANN, the time step is either fixed or required as an additional input to the model. Unless treated explicitly within the ANN architecture, the integration scheme contained within the ANN in such methods will naturally reflect the scheme used to recover the training data itself. Lastly, this approach is completely black-box in nature, and few constraints based on underlying physical relations (i.e. Arrhenius form) can be exploited (though some constraints, such as mass fraction conservation, can be enforced with the correct output layer activation function \cite{sharma2020deep}). A consequence of the black-box quality lies in ANN interpretability: if one deploys a trained ANN using this approach into a solver, by construction, assessing where and how the resulting model fails is difficult because of its large operation scope. 

The second category is to replace just the source term computation with a trained ANN that takes thermochemical state as input and outputs the corresponding source terms (ANN as a tabulation method) \cite{menon_ann,shivam_ftc,ranade_ann}. Here, the ANN serves as an approximation to a known nonlinear function. This can be seen as narrowing the scope of the ANN with respect to the first approach, as it eliminates the time-integrator role played by the neural network. Due to this elimination, the advantage here is that well-established GPU (or other ANN-based) integration techniques \cite{niemeyer2014,neural_ode} can still be utilized, and the role of the ANN becomes more transparent. However, the disadvantage is still in the prohibitive dependence on sampling an high-dimensional dynamical system to produce the training data. Techniques that rely on clustering subsets of the thermochemical state within the ANN \cite{rigopolous_ANN,shivam_ftc} have been attempted to reduce this dependence, though this adds significant computational complexity to the architecture and introduces additional assumptions to the procedure. Overall, both routes discussed above are at risk of overfitting to the configurations used to obtain the training data \cite{christo_ann}. In-situ training techniques \cite{pope_isat,isat_ann} can be used in the ANN setting in light of this and remains an open area of research, though it ambitiously relies on the in-situ training phase to be overall less expensive than the deployment phase.

The goal of this work concentrates on the second category, but adopts a different approach -- the exact equations for chemical kinetics are cast into an ANN-based form. More specifically, components of the source term evaluation are transformed into matrix-multiplication representations that can be interpreted as ANN layers. It will be shown that this “exact ANN” framework can be modified by utilizing trained ANNs as drop-in replacements for their exact form counterparts. Such replacements allow for direct control over the computational cost of individual components of the source term evaluation through their architectures, and because of this, additional speedup on the GPU can be extracted over the corresponding "exact ANN" form in some conditions. Furthermore, it will be shown that the training approach for these drop-in replacements a) does not require an intensive high-dimensional sampling procedures for data generation, and b) allows the user to retain certain physical constraints that drive the source term computation (i.e. the Arrhenius form), providing a method that extends to any configuration. Note that the main objective of this paper is not to use machine learning for developing models, but rather to represent chemical kinetics using the ANN framework. The end result is that little-to-no training is needed while preserving the GPU-friendly structure of the ANN formulation. 

The remainder of the paper proceeds as follows. In Sec.~\ref{sec:methodology}, the methodology for the ANN-inspired formulation is presented. In Sec.~\ref{sec:results}, the method is demonstrated using various simulations, and GPU speedup and saturation effects are discussed. Concluding remarks are provided in Sec.~\ref{sec:conclusion}. 

\section{Neural Network Interpretation of Kinetics}
\label{sec:methodology}
The main goal of this section is to recast the chemical kinetics equations into neural network interpretations that enable fast GPU execution through matrix multiplications. Though these reformulations are exact, they can be easily modified to include trained artificial neural networks (ANNs) to further enhance speedup. It will be shown that this form of ANN integration allows for extendable models that preserve underlying physical constraints attributed to the Arrhenius form and chemistry mechanism structure.

In the following, the quantities $N_C$, $N_S$ and $N_R$ denote the batch size (which can be interpreted as the number of reacting cells in a domain offloaded to the GPU), number of species, and number of reactions, respectively. Unless otherwise indicated, matrices are denoted by bold symbols ($\bf A$) and vectors by non-bold symbols. The scalar entry of matrix $\bf A$ in row $i$ and column $j$ is denoted ${\bf A}_{ij}$; similarly, the scalar $i$-th entry of vector $a$ is denoted $a_i$. Further, the quantities $i$, $j$, and $k$ index $N_C$, $N_R$, and $N_S$ respectively (i.e. $i = 1,\ldots,N_C$, $j = 1,\ldots, N_R$, and $k = 1,\ldots,N_S$). For the set of species $\{ \mathcal{S}_1, \ldots, \mathcal{S}_{N_S} \}$, a general chemical mechanism is represented as
\begin{equation}
    \sum_{k=1}^{N_S} {\boldsymbol{\nu}}'_{kj} \mathcal{S}_k \rightleftharpoons 
    \sum_{k=1}^{N_S} {\boldsymbol{\nu}}''_{kj} \mathcal{S}_k, \quad j = 1,\ldots,N_R,
\end{equation}
where $\boldsymbol{\nu}' \in \mathbb{R}^{N_S \times N_R}$ (resp. $\boldsymbol{\nu}''$) is the reactant (resp. product) stoichiometric coefficient matrix, and $\boldsymbol{\nu} = \boldsymbol{\nu}'' - \boldsymbol{\nu}'$.

\subsection{Species Production Rate}
The molar net production rate ($kmol/m^3 s$) for species $k$ in cell $i$ is 
\begin{equation}
    \label{eq:sourceterm}
    {\bf \Omega}_{ik} = \sum_{j=1}^{N_R} {\boldsymbol{\nu}}_{kj}  {\bf Q}_{net_{ij}}, 
\end{equation}
where ${\bf \Omega} \in \mathbb{R}^{N_C \times N_S}$ contains the source terms and ${\bf Q}_{net} \in \mathbb{R}^{N_C \times N_R}$ contains the net reaction rates. Note that Eq.~\ref{eq:sourceterm} can be expressed concisely through the matrix multiplication ${\bf \Omega} = {\bf Q}_{net} \boldsymbol{\nu}^T$. The complexity comes from the net reaction rate, which is expressed as 
\begin{equation}
    \label{eq:netrate}
    {\bf Q}_{net_{ij}} = {\bf Q}_{f_{ij}} - {\bf Q}_{r_{ij}} = {\bf K}_{{f}_{ij}} \prod_{k=1}^{N_S} {\bf C}_{ik}^{\nu_{kj}'} - {\bf K}_{{r}_{ij}} \prod_{k=1}^{N_S} {\bf C}_{ik}^{\nu_{kj}''}.
\end{equation}

Above, ${\bf Q}_{f}$ and ${\bf Q}_{r} \in \mathbb{R}^{N_C \times N_R}$ are the forward and reverse reaction rate matrices respectively, ${\bf K}_{f}$ and ${\bf K}_{r} \in \mathbb{R}^{N_C \times N_R}$ are the forward and reverse rate constants respectively, and ${\bf C} \in \mathbb{R}^{N_C \times N_S}$ contains the species molar concentrations. Since ${\bf Q}_{f}$ and ${\bf Q}_{r}$ are non-negative, Eq.~\ref{eq:netrate} can be interpreted as a summation of two ANN layers by enabling matrix multiplications in the logarithm space:
\begin{equation}
    \label{eq:netrate_log}
    {\bf Q}_{net} = 
    \exp\big(\log({\bf C}) \boldsymbol{\nu}' + \log({\bf K}_f) \big) - 
    \exp\big(\log({\bf C}) \boldsymbol{\nu}'' + \log({\bf K}_r) \big).
\end{equation}

It can be seen through Eq.~\ref{eq:netrate_log} that the forward and reverse contributions are ANN layers with exponential activation functions, where the input is the logarithm of the concentration matrix $\bf C$, the weight matrices are known stoichiometric coefficients $\boldsymbol{\nu}'$ and $\boldsymbol{\nu}''$, and the biases are the logarithms of rate constants  ${\bf K}_f$ and ${\bf K}_r$. 

\begin{figure}
    \centering
    \includegraphics[width=\columnwidth]{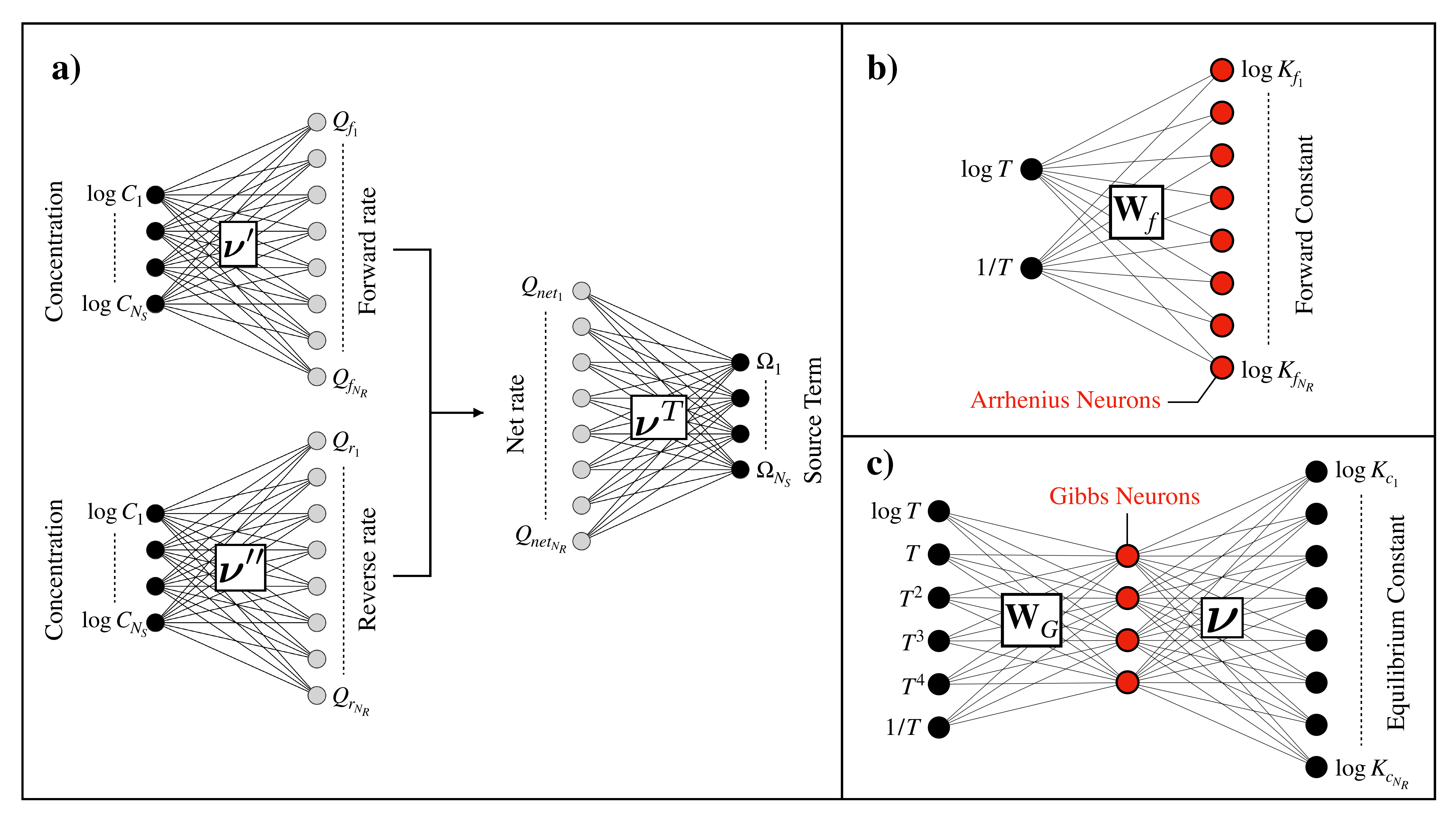}
    \caption{Illustrations ANN-based formulations for $N_C=1$, $N_S=4$, and $N_R=8$. Since $N_C=1$, input/outputs are vectors and cell indices are ignored. Bias terms not shown for clarity. \textbf{a)} Schematic of Eqs.~\ref{eq:sourceterm} and \ref{eq:netrate_log}. Exponential activation functions are used to produce forward/reverse rates. \textbf{b)} Schematic of Arrhenius layer for forward rate constant (Eq.~\ref{eq:log_kf}). \textbf{c)} Schematic of Gibbs layer equilibrium constant (Eq.~\ref{eq:gibbs_ann}).}
    \label{fig:ann_sourceterm}
\end{figure}

Figure~\ref{fig:ann_sourceterm}a summarizes the above formulation (Eqs.~\ref{eq:sourceterm} and \ref{eq:netrate_log}) through an ANN architecture. Note that the leading matrix dimension of all input and output variables, which constitutes the batch size in the forward pass, is $N_C$. This allows for efficient threading and fast execution in high fidelity settings, assuming optimized linear algebra libraries (such as cuBLAS) are utilized by the user. The remaining task, described below, is to obtain the rate constants ${\bf K}_f$ and ${\bf K}_r$.

\subsection{Rate Constants}
\textbf{Forward Rate Constant:}

\noindent
The forward rate constant ${\bf K}_f \in \mathbb{R}^{N_C \times N_R}$ is given by the Arrhenius expression
\begin{equation}
    {\bf K}_{f_{ij}} = {A}_j T_i^{\beta_j} \exp \left(-\frac{E_j}{RT_i}\right), 
\end{equation}
where $A$, $\beta$, and $E$ are vectors each of size $N_R$ containing pre-exponential factors, temperature exponents, and activation energies respectively for the elementary reactions. These Arrhenius parameters are known to the user through the mechanism files. The natural logarithm of the forward rate (required in Eq.~\ref{eq:netrate_log}) usefully yields a form that can also be interpreted as a linear ANN layer,
\begin{equation}
    \label{eq:log_kf}
    \log({\bf K}_f) = {\bf X}_f {\bf W}_f + B_f, \text{ where}
\end{equation}
\begin{gather*}
\begin{aligned}
    {{\bf X}_f } = \begin{bmatrix}
    \log T_1 & 1/T_1\\
    \log T_2 & 1/T_2\\
    \vdots & \vdots \\
    \log T_{N_C} & 1/T_{N_C}
    \end{bmatrix}, \quad
    {\bf W}_f = \begin{bmatrix}
    \beta_1 & \cdots & \beta_{N_R}\\
    -E_1/R & \cdots & -E_{N_R}/R
    \end{bmatrix}, \quad
    {B}_f = \begin{bmatrix}
    \log A_1\\
    \log A_2\\
    \vdots\\
    \log A_{N_R}
    \end{bmatrix}^T.
\end{aligned}
\end{gather*}

In Eq.~\ref{eq:log_kf}, ${\bf X}_f \in \mathbb{R}^{N_C \times 2}$ is the temperature-dependent input, ${\bf W}_f \in \mathbb{R}^{2 \times N_R}$ is a weight matrix consisting of temperature exponents and activation energies, and $B_f \in \mathbb{R}^{1 \times N_R}$ is a bias term of pre-exponential factors. In this sense, each row of the layer output $\log({\bf K}_f)$ can be interpreted as a set of $N_R$ Arrhenius neurons. 


\vskip 0.2in
\noindent
\textbf{Reverse Rate Constant:}

\noindent
The reverse rate constant ${\bf K}_r \in \mathbb{R}^{N_C \times N_R}$ is given by 
\begin{equation}
    {\bf K}_{r_{ij}} = {\bf K}_{f_{ij}} / {\bf K}_{c_{ij}}, 
\end{equation}
where ${\bf K}_c \in \mathbb{R}^{N_C \times N_R}$ contains the equilibrium rate constants. Since the expression for $\log ({\bf K}_f)$ is provided through the Arrhenius neurons (Eq.~\ref{eq:log_kf}), the task of determining $\log ({\bf K}_r)$ required in Eq.~\ref{eq:netrate_log} is accomplished by considering only $\log ({\bf K}_c)$. 

The equilibrium constant for cell $i$ and reaction $j$ is \cite{poinsot_book} 
\begin{equation}
    \label{eq:kc}
    {\bf K}_{c_{ij}} = \left(\frac{p_{ref}}{R T_i} \right)^{\sum_k {\boldsymbol{\nu}_{kj}}} \exp \left( \frac{\Delta S_j(T_i)}{R} - \frac{\Delta H_j (T_i)}{R T_i} \right),
\end{equation}
where $\Delta S_j$ and $\Delta H_j$ are changes in entropy and enthalpy for reaction $j$, and $p_{ref}$ is the reference pressure (1 bar). The logarithm of Eq.~\ref{eq:kc} yields
\begin{equation}
    \label{eq:log_kc}
    \log( {\bf K}_{c_{ij}}) = \sum_{k=1}^{N_S} \boldsymbol{\nu}_{jk} \left( -{\bf G}_{ik} +  \frac{p_{ref}}{R T_i} \right),
\end{equation}
where ${\bf G} \in \mathbb{R}^{N_C \times N_S}$ is the nondimensional Gibbs free energy matrix (hereafter referred to as the Gibbs matrix) obtained from the nondimensional enthalpy (${\bf H} \in \mathbb{R}^{N_C \times N_S}$) and entropy (${\bf S} \in \mathbb{R}^{N_C \times N_S}$) matrices. Each entry in the Gibbs matrix is determined from NASA polynomials which provide species enthalpy and entropy as tabulated functions of temperature. The result can be expressed as a matrix multiplication
\begin{equation}
    \label{eq:gibbs}
    {\bf G} = {\bf H} - {\bf S} = {\bf X}_G {\bf W}_G + B_G, \text{ where}
\end{equation}
\begin{gather*}
\begin{aligned}
    {{\bf X}_G } = \begin{bmatrix}
    \log T_1 & T_1 & T_1^2 & T_1^3 & T_1^4 & 1/T_1\\
    \log T_2 & T_2 & T_2^2 & T_2^3 & T_2^4 & 1/T_2\\
    \vdots & \vdots & \vdots & \vdots & \vdots & \vdots\\
    \log T_{N_C} & T_{N_C} & T_{N_C}^2 & T_{N_C}^3 & T_{N_C}^4 & 1/T_{N_C}
    \end{bmatrix}, \quad
    {{\bf W}_G } = \begin{bmatrix}
    \alpha_{1,1} & \hdots & \alpha_{1,N_S}\\
    \alpha_{2,1} & \hdots & \alpha_{2,N_S}\\
    \vdots & \ddots & \vdots\\
    \alpha_{6,1} & \hdots & \alpha_{6,N_S}
    \end{bmatrix}, \quad
    { B_G } = \begin{bmatrix}
    \alpha_{7,1}\\
    \alpha_{7,2}\\
    \vdots\\
    \alpha_{7,N_S}
    \end{bmatrix}^T.
\end{aligned}
\end{gather*}

In Eq.~\ref{eq:gibbs}, ${\bf X}_G \in \mathbb{R}^{N_C \times 6}$ is the input consisting of various functions of temperature and ${\boldsymbol{\alpha}} \in \mathbb{R}^{7 \times N_S}$ is a matrix of polynomial coefficients; the first 6 rows of ${\boldsymbol{\alpha}}$ is the weight matrix ${\bf W}_G$ and the last row is the bias $B_G$. Note that, though not shown in Eq.~\ref{eq:gibbs} for conciseness, the quantities in ${\boldsymbol{\alpha}}$ (and in turn ${\bf W}_G$ and $B_G$) are also functions of the cell temperature $T_i$ and the species index. This is because the species polynomial coefficients change based on a cutoff temperature (usually 1000 K). 

Inserting Eq.~\ref{eq:gibbs} into Eq.~\ref{eq:log_kc} gives
\begin{equation}
    \label{eq:gibbs_ann}
    \log({\bf K}_c) = -({\bf X}_G {\bf W}_G + B_G) \boldsymbol{\nu},
\end{equation}
where the standard concentration term $p_{ref}/RT_i$ has been integrated into the bias $B_G$. Equation~\ref{eq:gibbs_ann} can be interpreted as a linear two-layer ANN. The parameters of the first layer (the Gibbs layer) are the temperature-dependent ${\bf W}_G$ and $B_G$, and those of the second layer are the net stoichiometric coefficients $\boldsymbol{\nu}$. The intermediary neurons (i.e. hidden layer neurons) here are referred to as the Gibbs neurons. 

Illustrations of the both forward and equilibrium rate constant formulations as neural network-inspired architectures are shown in Fig.~\ref{fig:ann_sourceterm}b and c, with Arrhenius and Gibbs neurons highlighted. 

\subsection{Integration of Approximate Artificial Neural Networks}
\label{sec:approx_ann}
Thus far, the ANN-inspired reformulations are exact and can by themselves be implemented on GPUs with efficient linear algebra libraries -- we refer to these as exact ANNs. However, additional computational efficiency can be provided by utilizing approximate ANNs as drop-in replacements for the exact ANN architectures described in Fig.~\ref{fig:ann_sourceterm}. At the cost of accuracy, such replacements allow for direct control over computational cost through the ANN architecture. The end goal is that the execution time of the approximate ANN should be faster than the exact ANN counterpart. 

Though many pathways to this end are available, here, the replacement of the exact ANN for the logarithm of the equilibrium rate constant (Fig.~\ref{fig:ann_sourceterm}c) will be explored. By narrowing the approximate ANN scope to the equilibrium constant alone, a) sampling an $N_S$-dimensional phase space to develop a training dataset is not required, and b) known physical constraints to recover the source term, such as the relationship between concentrations and net reaction rate (Eq.~\ref{eq:netrate_log}), and Arrhenius forms (Eq.~\ref{eq:log_kf}), are preserved. Additionally, the exact ANN architecture in Fig.~\ref{fig:ann_sourceterm}c is complex enough to warrant a reduction based on an approximate ANN (there is more than one layer, which is not the case for the forward rate constant architecture).

In general, an ANN layer takes the following form: 
\begin{equation}
    {\bf X}_{l+1} = \sigma_l( {\bf X}_{l} {\bf W}_l + B_l ), \quad l = 0,\ldots,N_L,
\end{equation}
where ${\bf X}_l$ is the layer input, ${\bf X}_{l+1}$ is the layer output, ${\bf W}_l$ is the weight matrix, $B_l$ is the bias vector, $\sigma$ is an activation function, and $N_L$ is the total number of hidden layers. Note that unlike in the above sections, the parameters (weights/biases) are assumed unknown in this setting and are found through a training procedure. As with the exact formulations, the leading dimension (batch size) for these input and output matrices is $N_C$. Here, to simplify analysis, for a given $N_L$, the hidden layer dimension $N_H$ (or number of neurons per hidden layer) is fixed.

\begin{figure}
    \centering
    \includegraphics[width=\columnwidth]{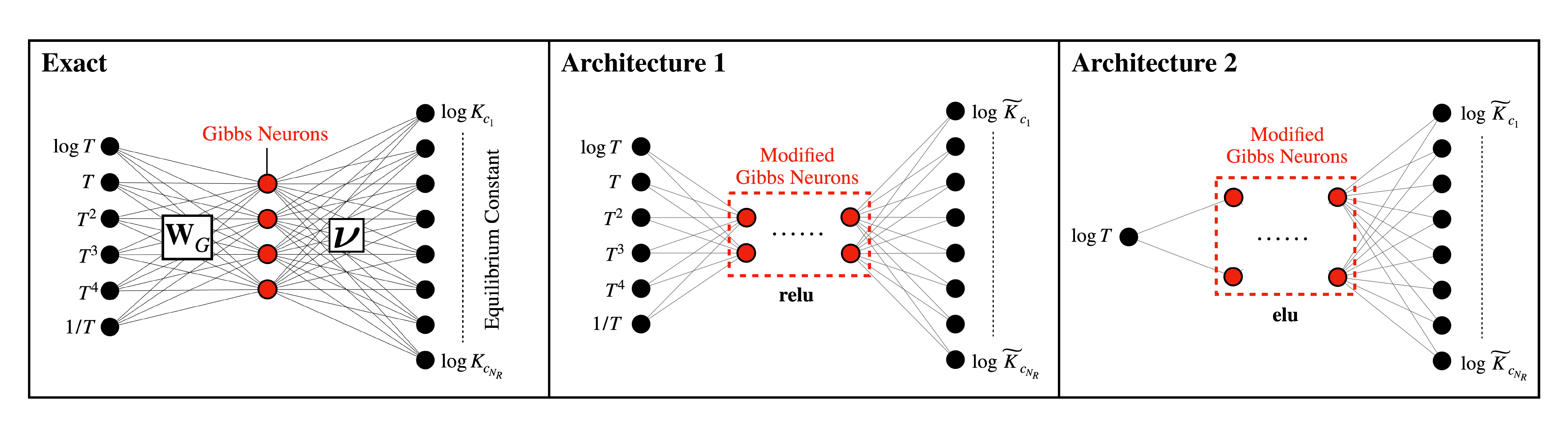}
    \caption{Illustrations of ANN representations for $\log( \bf{K_c} )$. (Left) Exact ANN (same as Fig.~\ref{fig:ann_sourceterm}c). (Middle) Approximate ANN, Architecture 1. (Right) Approximate ANN, Architecture 2. The number of modified Gibbs neurons changes through $N_H$ and $N_L$ ($N_H=2$ here for illustrative purposes).}
    \label{fig:trained_ann}
\end{figure}

The ANN input is ${\bf X}_{0} \in \mathbb{R}^{N_C \times N_{in}}$ and the output is ${\bf X}_{N_L+1} \in \mathbb{R}^{N_C \times N_R} = \log(\widetilde{{\bf K}_c})$. The ANN is trained such that $\log(\widetilde{{\bf K}_c}) \approx \log({\bf K}_c)$. The only restrictions are that the input features are functions of temperature and the output dimensionality is $N_R$. To highlight key points, two examples of approximate ANN architectures are shown in Fig.~\ref{fig:trained_ann}. Architecture 1 uses the same input features as the exact ANN but allows for variation $N_L$ and $N_H$ (referred to as the modified Gibbs neurons). Architecture 2 is similar but only utilizes one input feature, namely $\log(T)$. 

Nonlinearity is imposed in both architectures through the activation functions $\sigma_l$. In Architecture 1, since several functions of temperature are already included in the input, a simple rectified linear unit (relu) activation function can be used:
\begin{equation}
    \forall x \in \mathbb{R},\quad \sigma_l(x) = \text{relu}(x) = \max(0,x).
\end{equation}

On the other hand, since Architecture 2 utilizes only $\log(T)$ in the input layer, the more expensive exponential linear unit (elu) activation can be used \cite{elu} to allow the model to extract dependence on powers of $T$ as needed during the training process:
\begin{equation}
    \forall x \in \mathbb{R},\quad 
    \sigma_l(x) = \text{elu}(x) = 
        \begin{cases}
                x &\text{if } x \geq 0,\\
                e^x - 1 &\text{if } x < 0.
        \end{cases}
\end{equation}

In the above scenario, it is reasonable to expect that the computational advantage offered by the smaller input size of Architecture 2 is offset by the more expensive activation function. In light of this, Architecture 2 can be modified to use the relu activation, though this lessens the expressive power of the ANN. Although several other candidate architectures can be created, in the results below, we demonstrate the approximate ANN performance using only Architecture 2 for brevity, as this architecture consists of a less complex input. Overall trends discussed throughout this work are applicable to both (and more) architecture types. As an aside, in some cases it is advantageous if the architecture is modified to instead operate on scaled versions of the inputs and outputs, which can assist in weight convergence during the training phase. In this work, a standardization operation (subtraction of training set mean and normalization by training set standard deviation) is used for scaling, though many other scaling strategies are viable. 

\subsection{Additional Comments}
\textbf{Three-body reactions:} To handle three-body reactions, the quantity $\log({\bf M})$ can be added to $\log({\bf K}_f)$, where ${\bf M} \in \mathbb{R}^{N_C \times N_R}$ is a matrix of third-body concentrations for each reaction (${\bf M}_{ij}$ is $1$ if reaction $j$ does not include a third body). The entries in ${\bf M}$ can be obtained through the matrix multiplication ${\bf M} = {\bf C} {\bf E}$, where ${\bf E} \in \mathbb{R}^{N_S \times N_R}$ is a matrix of third-body efficiency factors. 

\vskip 0.2in
\noindent
\textbf{Falloff reactions:} Falloff reactions are treated separately from standard reactions -- although the Arrhenius rate constants used to compute the non-dimensional reduced pressure can be treated by the same Arrhenius layer described in Fig.~\ref{fig:ann_sourceterm}b, falloff functions such as that of Troe (if they exist in the mechanism) are handled separately on the GPU in a non-matrix fashion. A useful extension of the method could be to capture both standard and Troe falloff functions in a single forward pass using an approximate ANN for the forward rate constant. 

\vskip 0.2in
\noindent
\textbf{Irreversible reactions:} Treatment of irreversible reactions is handled by appending an indicator function encoding reaction reversibility to the reverse reaction rate term in Eq.~\ref{eq:netrate}.

\section{Results}
\label{sec:results}
The objective below is to first demonstrate the feasibility of the approximate ANN replacement (Fig.~\ref{fig:trained_ann}, Architecture 2) for the equilibrium rate constant as described in Sec.~\ref{sec:approx_ann}, and to then assess the GPU-based performance of the method (Sec.~\ref{sec:gpu}). Table~\ref{tab:mechanisms} details the three mechanisms (referred to as mechanisms A, B, and C) of increasing complexity used throughout this section. 

\begin{table}[]
    \centering
    \resizebox{\textwidth}{!}{%
    \begin{tabular}{c|c|c|c|c|c|c|c|}
                  & \textbf{Name} & \textbf{Description} & \textbf{Species} & \textbf{Reactions} & \textbf{HF-ANN} & \textbf{LF-ANN} & \textbf{T [K]} \\
       \textbf{Mech. A} & Mueller et al. \cite{hon_mueller} & $\text{H}_2$/Air & 9 & 21 & 3/32 & 1/8 & 200 -- 6000 \\
       \textbf{Mech. B} & FFCMy-12 \cite{wang_methane,smith_methane} & $\text{CH}_4$/$\text{O}_2$ & 12 & 38 & 3/32 & 2/8 & 200 -- 6000 \\
       \textbf{Mech. C} & UCSD \cite{ucsd} & $\text{H}_2$/Air  & 57 & 168 & 3/32 & 1/8 & 300 -- 5000
    \end{tabular}}
    \caption{Details of chemistry mechanisms used throughout Sec.~\ref{sec:results}. Last three columns show high-fidelity (HF) ANN architecture used in predictions in Sec.~\ref{sec:prediction}, low-fidelity (LF) architecture used in predictions in Sec.~\ref{sec:prediction}, and temperature range used to generate the training data, respectively. For ANN columns, notation "X/Y" refers to ANN with X hidden layers and Y neurons per hidden layer.}
    \label{tab:mechanisms}
\end{table}

\subsection{ANN Demonstration}
\label{sec:ann_demo}
\subsubsection{Training results:}
\label{sec:training}
As per Fig.~\ref{fig:trained_ann}, the goal of the approximate ANN is to recover the logarithm of the equilibrium rate constant for all reactions in the mechanism, $\log( {\bf K}_c )$. As such, the training data is obtained by sweeping through a range of temperatures (functions of which supply the input) and, for each temperature, recovering the exact $N_R$-dimensional vector of equilibrium rate constants (which supplies the target). Then, standard supervised techniques can be used to train the neural network with a specified loss function. The mean-squared error (MSE) loss on the scaled (standardized) logarithm of the equilibrium constant is used here. Note that this approach usefully eliminates an involved high-dimensional sampling procedure for obtaining the training data -- temperature is trivially sampled from a range that is usually known a-priori. For each mechanism, 1 million temperature / equilibrium constant pairs were sampled within the ranges specified in Tab.~\ref{tab:mechanisms} (i.e. $N_C=10^6$ in the training phase). ANNs were trained with the PyTorch library \cite{pytorch}. 

\begin{figure}
    \centering
    \includegraphics[width=0.7\columnwidth]{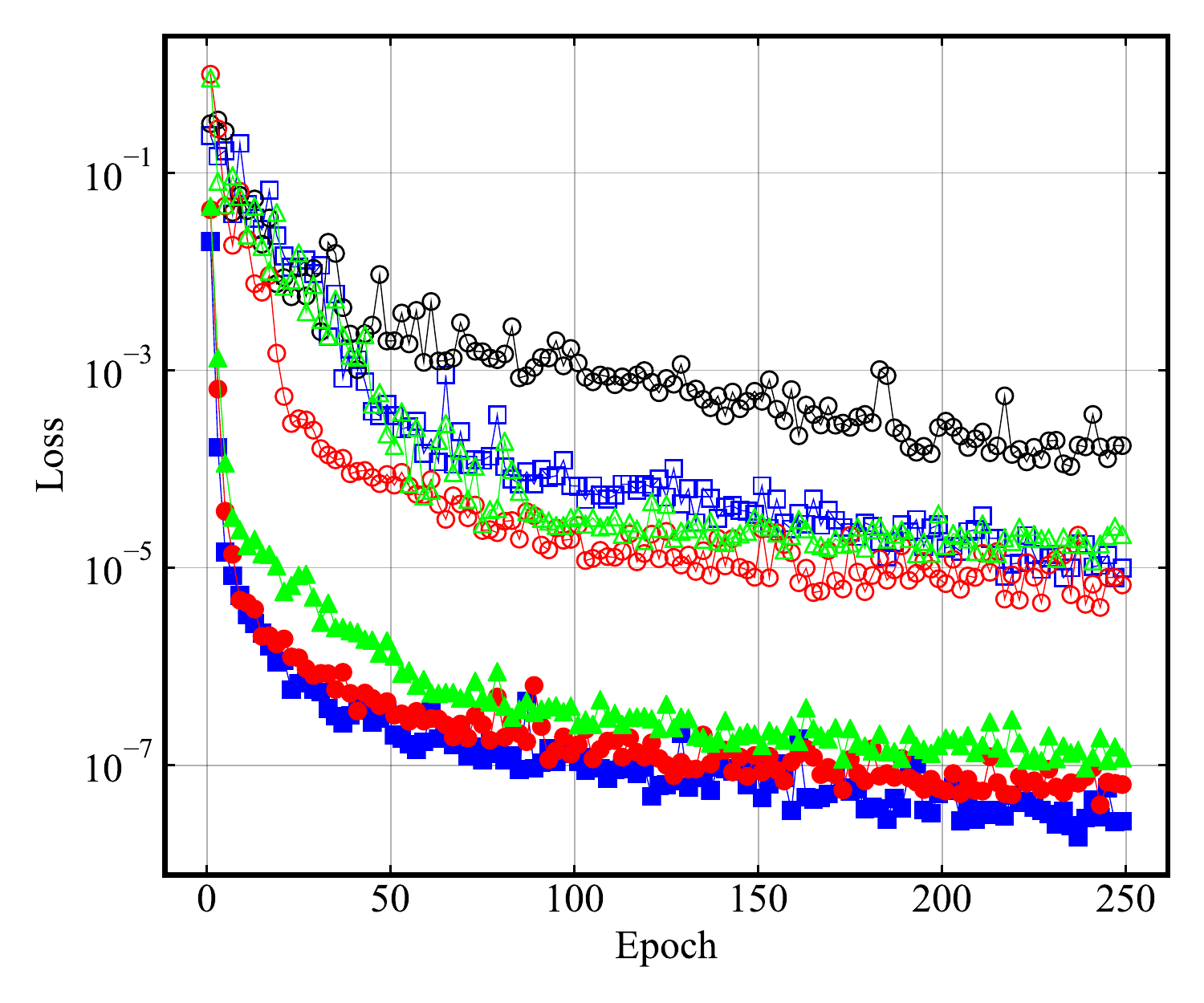}
    \caption{
    ANN training histories. Loss is MSE normalized by the maximum MSE over all models. Notation "X/Y" refers to an ANN with X hidden layers and Y neurons per hidden layer. \textbf{Mechanism A}: 1/8 ANN (\mythickbarredsquare{blue}{white}), 3/32 ANN (\mythickbarredsquare{blue}{blue}). \textbf{Mechanism B:} 1/8 ANN (\mythickbarredcircle{black}{white}), 2/8 ANN (\mythickbarredcircle{red}{white}), 3/32 ANN (\mythickbarredcircle{red}{red}). \textbf{Mechanism C:} 1/8 ANN (\mythickbarredtriangle{green}{white}), 3/32 ANN (\mythickbarredtriangle{green}{green}).
    }
    \label{fig:training}
\end{figure}

Training results for the three mechanisms are shown in Fig.~\ref{fig:training} in the form of loss function history versus training iteration (epoch). For all models and mechanisms shown, training parameters (batch size, learning rate, total epochs, optimization method, etc.) were fixed to allow for more direct comparisons. Two classes of models are considered: low-fidelity (LF) and high-fidelity (HF) ANNs. The architecture for the HF-ANNs (3 hidden layers, 32 neurons per layer) is constructed such that its parameter space is much larger relative to the LF-ANN counterparts (1-to-2 hidden layers, 8 neurons per layer). The consideration of ANNs with varying complexity is important in the context of speedup, to be explored in Sec.~\ref{sec:gpu}. In the discussion below, the notation X/Y is used to concisely refer to an ANN with X hidden layers and Y neurons per hidden layer.

Figure~\ref{fig:training} shows that the HF-ANNs (the 3/32 ANNs) approach MSE values that increase slightly with increasing mechanism complexity. Despite this, the convergence point for all three HF-ANNs occur at nearly the same order of magnitude. In other words, for the ANN architectures considered, no significant decrease in training loss is seen when moving from Mechanism A to C. This is particularly impressive because Mechanism C is much more complex than A. 

As expected, the LF-ANNs approach MSE values much higher than the HF-ANN counterparts. Surprisingly, despite the vast difference in complexity, the 1/8 ANNs for Mechanisms A and C approach the same MSE. Interestingly, the converged MSE for the 1/8 ANN for Mechanism B is roughly an order of magnitude lower than both Mechanisms A and C. Figure~\ref{fig:training} shows that a jump from $1/8$ to $2/8$ (addition of one hidden layer) in the LF-ANN architecture is required for Mechanism B to converge to a loss similar to the $1/8$ ANNs for Mechanisms A and C. This highlights an important facet of chemical mechanism complexity: although Mechanism B contains less species and reactions than C, higher order functions of temperature are required in the ANN estimation to recover the equilibrium constants at similar levels of accuracy. 

\subsubsection{Predictions:}
\label{sec:prediction}

The remainder of this section outlines how errors in the $\log({{\bf K}_c})$ computation (Fig.~\ref{fig:training}) translate to errors in mass fraction predictions in \textit{a-posteriori} simulation settings. This involves replacing the exact equilibrium rate constants with the approximate ANN outputs during the source term computation used throughout the simulation. Both HF and LF-ANN models for each mechanism are considered here; architectures are supplied in Tab.~\ref{tab:mechanisms}. Note that in light the discussion in Sec.~\ref{sec:training}, the LF-ANN architecture for Mechanism B for the analysis below contains two hidden layers instead of one. Two simulation scenarios are considered: zero-dimensional (0-D) auto-ignition and one-dimensional (1-D) channel detonation.

\vskip 0.2in
\noindent
\textbf{0-D Ignition:}

\noindent
Auto-ignition in a constant pressure reactor was simulated for the three mechanisms using both exact and ANN-based formulations. All simulations were performed using a forward Euler integration scheme at a constant time step of 1e-10s. Only one initial condition per mechanism is shown here for conciseness.

\begin{figure}
    \centering
    \includegraphics[width=0.65\columnwidth]{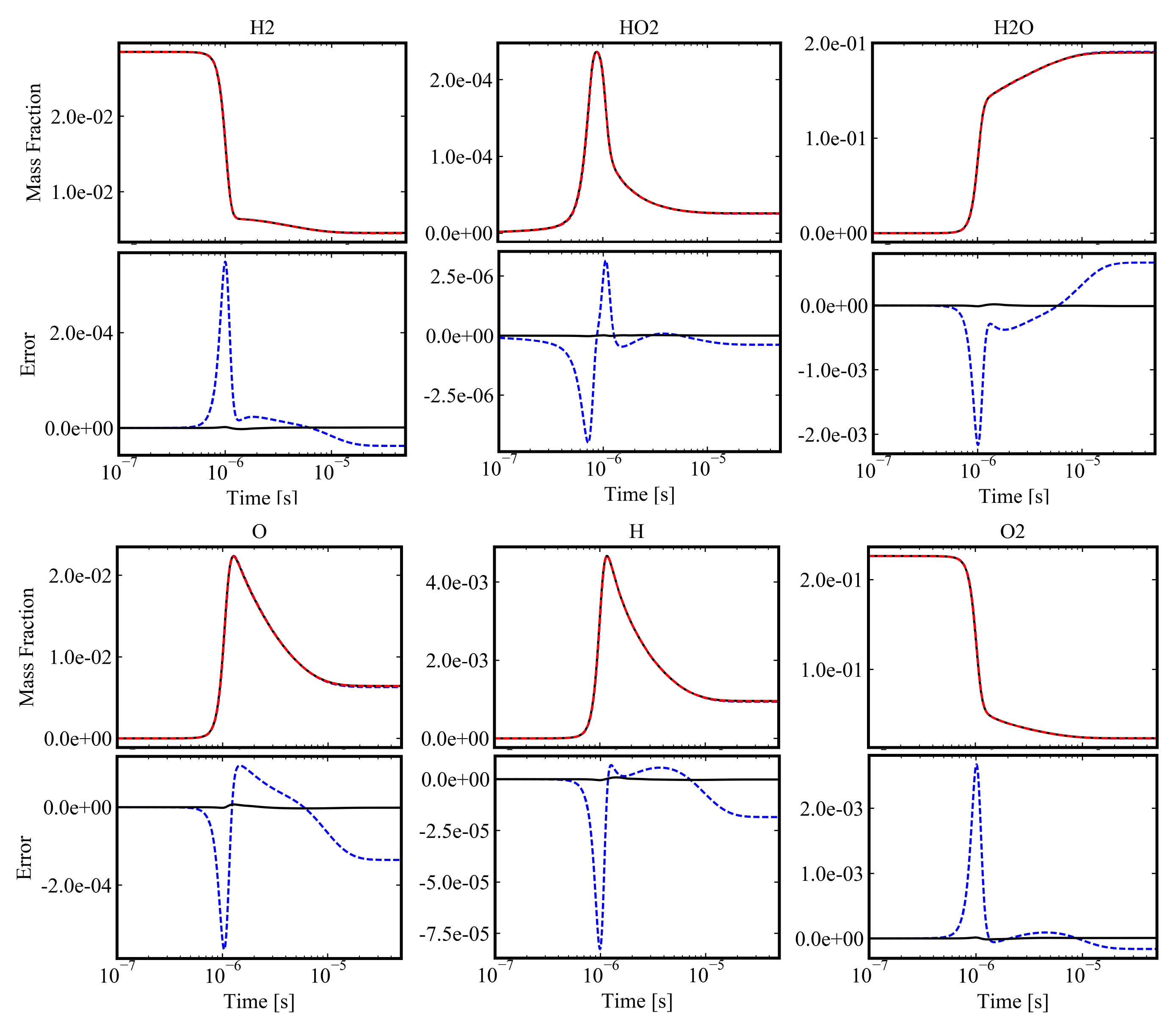}
    \caption{Mechanism A species profiles for exact (\mythickdashedline{red}), LF-ANN (\mythickdashedline{blue}), and HF-ANN (\mythickline{black}) based 0-D ignition simulations (hydrogen/air). Time evolution of errors computed as difference between exact and ANN profiles are shown below each mass fraction plot. Initial conditions: T = 1800 K, P = 5 atm, $\phi=1$. Air mixture includes $\text{H}_2$, $\text{O}_2$, and $\text{N}_2$.}
    \label{fig:zeroD_A}
\end{figure}

\begin{figure}
    \centering
    \includegraphics[width=0.65\columnwidth]{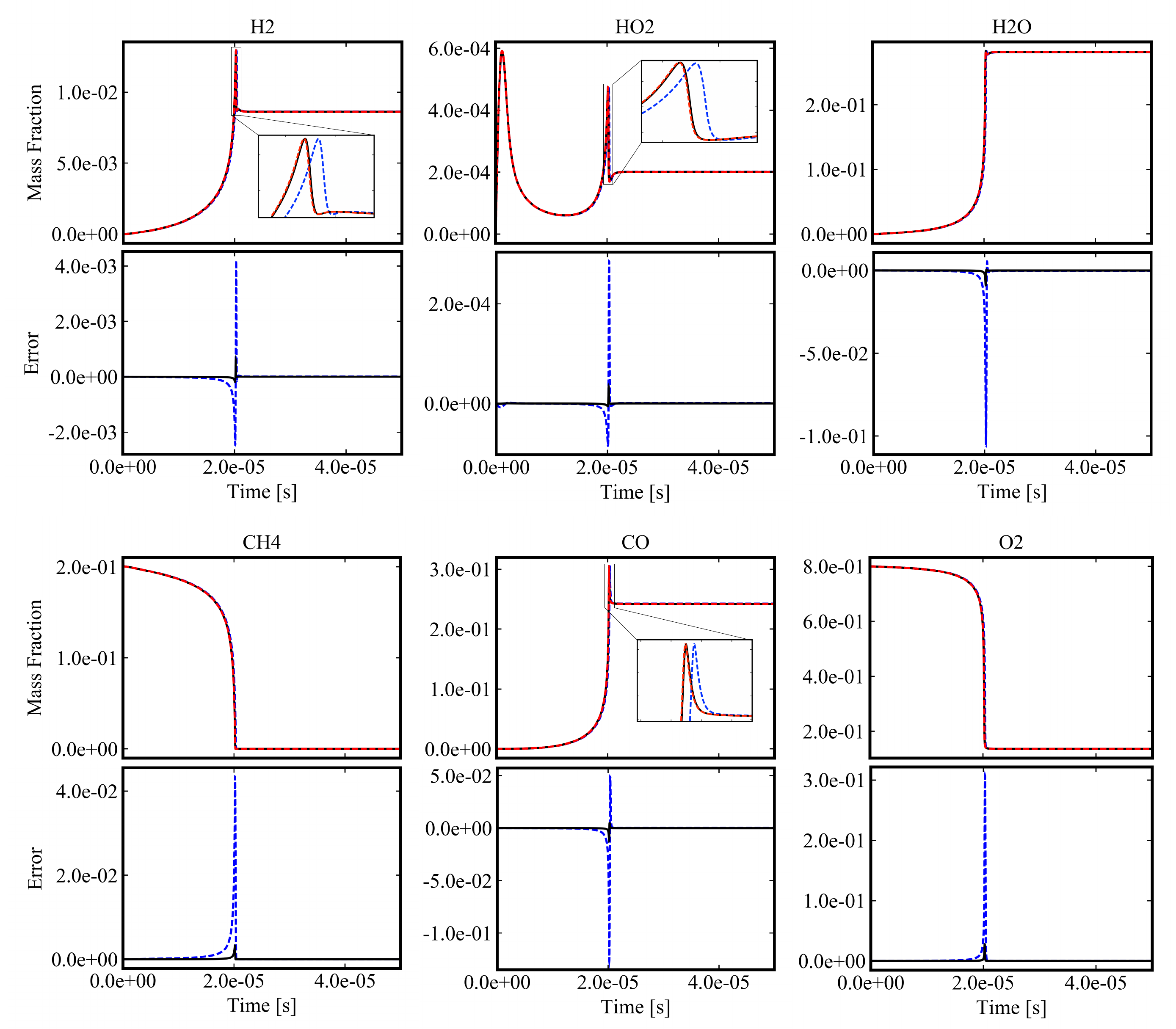}
    \caption{Mechanism B species profiles for exact (\mythickdashedline{red}), LF-ANN (\mythickdashedline{blue}), and HF-ANN (\mythickline{black}) based 0-D ignition simulations (methane/oxygen). Time evolution of errors computed as difference between exact and ANN profiles are shown below each mass fraction plot. Initial conditions: T = 1800 K, P = 5 atm, $\phi=1$.}
    \label{fig:zeroD_B}
\end{figure}

\begin{figure}
    \centering
    \includegraphics[width=0.65\columnwidth]{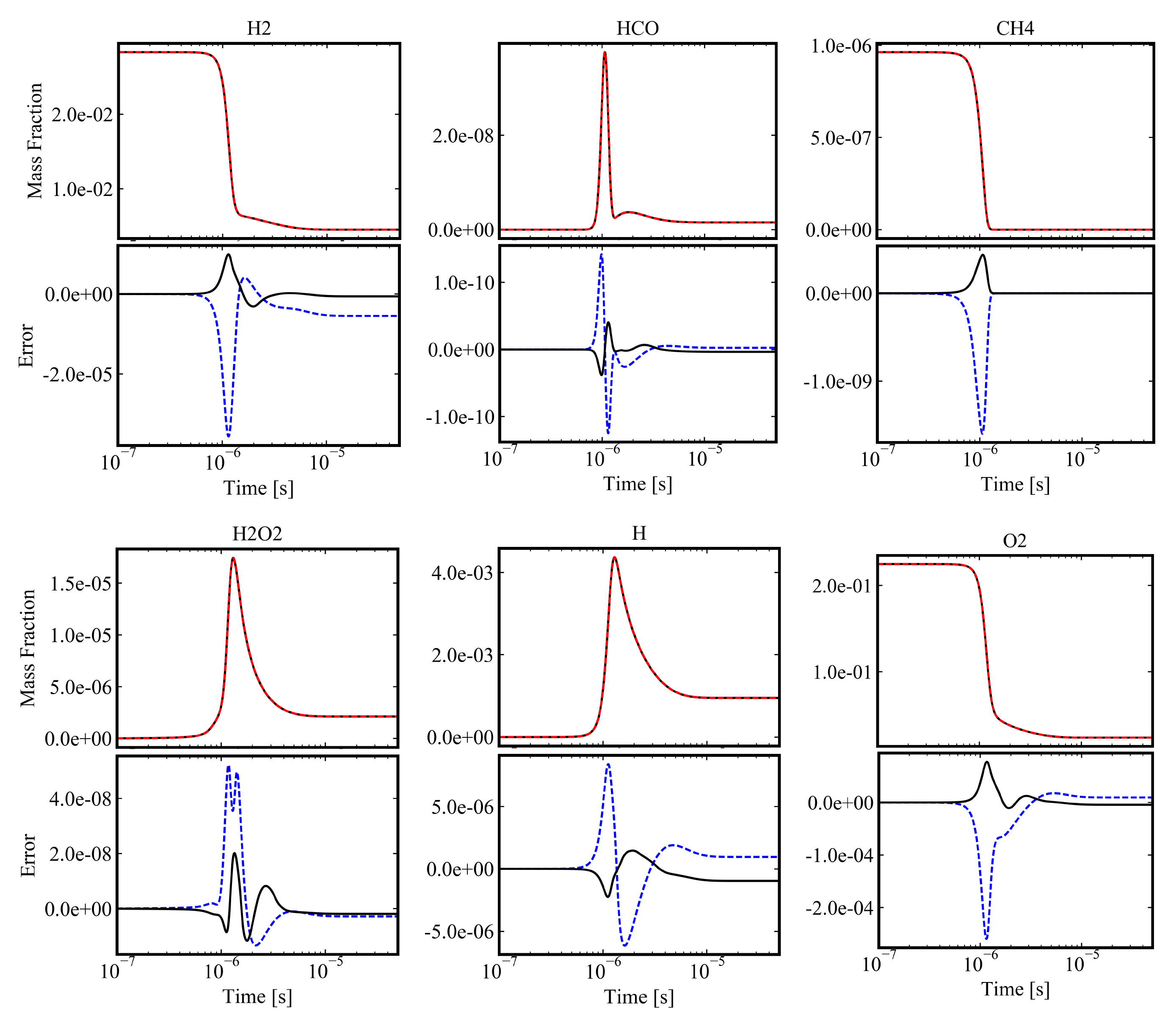}
    \caption{Mechanism C species profiles for exact (\mythickdashedline{red}), LF-ANN (\mythickdashedline{blue}), and HF-ANN (\mythickline{black}) based 0-D ignition simulations (hydrogen/air). Time evolution of errors computed as difference between exact and ANN profiles are shown below each mass fraction plot. Initial conditions: T = 1800 K, P = 5 atm, $\phi=1$. Air mixture includes $\text{H}_2$, $\text{O}_2$, $\text{N}_2$, Ar, He, $\text{CO}_2$, and $\text{CH}_4$.}
    \label{fig:zeroD_C}
\end{figure}

The time evolution of mass fraction and error for a subset of species are shown in Figs.~\ref{fig:zeroD_A}, \ref{fig:zeroD_B}, and \ref{fig:zeroD_C} for Mechanisms A, B and C respectively. The fuel and oxidizer used for Mechanisms A and C are hydrogen/air, whereas those for B are methane/oxygen. For all three mechanisms, the mass fraction profiles show that both the LF and HF-ANN based simulations are nearly indistinguishable from the exact (Cantera) counterpart. However, the error profiles expectedly reveal that the LF-ANN simulation observes significantly higher errors than the HF-ANN simulation, with peaks occurring near the ignition point. For Mechanisms A and C (Figs.~\ref{fig:zeroD_A} and \ref{fig:zeroD_C}), the highest observed LF-ANN errors occur at values of roughly two orders of magnitude lower than the respective mass fraction values. Surprisingly, the errors profiles for the LF-ANNs in Mechanism C (the mechanism with highest species and reaction count) are noticeably lower across the board -- this could likely be due to the reduced temperature range with which its training data was collected (see Tab.~\ref{tab:mechanisms}). 

Mechanism B (Fig.~\ref{fig:zeroD_B}) LF-ANN errors show peaks near or at the same order of the respective mass fraction values. These spikes in error are explained by the slightly offset peaks in the LF-ANN mass fractions. Since the ignition timescale is very small, this delay in ignition time (indicated by the insets in Fig.~\ref{fig:zeroD_B}) produces a very high mass fraction errors at the same time instant. This LF-ANN effect is not observed to the same degree for Mechanisms A and C, and is alleviated in Mechanism B by the HF-ANN. Despite this, the structure of the LF-ANN species profiles in Mechanism B are very close to the exact simulations; the relative error in LF-ANN ignition time shown in the insets in Fig.~\ref{fig:zeroD_B} is less than 1\%. 

These results ultimately show that a) despite higher MSE values observed in the training phase, the LF-ANN based approximations still produce near-exact mass fraction profiles, and b) one can correlate in confidence a drop in MSE in the context of Fig.~\ref{fig:training} with a drop in errors in species time evolution profiles for a given mechanism. However, the degree to which this error is reduced is not necessarily consistent across different mechanisms. An important result is that the LF-ANN maintains high accuracy even for Mechanism C; this is somewhat counter-intuitive because the LF-ANN provides significantly higher complexity reduction for Mechanism C than for the others. 

\vskip 0.2in
\noindent
\textbf{1-D Channel Detonation:}

\noindent
The 1-D detonation simulations were conducted with the OpenFOAM-based solver UMdetFOAM \cite{takuma_pci,jpc_takuma,prakash_pci}, which solves the governing equations of fluid flow consisting of mass, momentum, energy, and species conservation equations. UMdetFOAM is a compressible flow solver which contains shock-capturing numerics using the Monotonic Upwind Scheme for Conservation Laws (MUSCL)-based Harten-Lax-van Leer-Contact (HLLC) scheme \cite{asme_takuma,jpc_takuma}, a second-order Runge-Kutta temporal discretization with minimal dissipation \cite{malik-openfoampaper}, and the Kurganov, Noelle, and Petrova (KNP) scheme \cite{OF_rhocentral} for diffusion terms. For chemical reactions, the package Cantera \cite{goodwin2012cantera} is utilized in the purely CPU-based solver, whereas the GPU-offloaded UMdetFOAM uses the methodology outlined in Sec.~\ref{sec:methodology} implemented in the CUDA and cuBLAS environments. This code has been extensively validated using experiments of detonation-containing flows \cite{asme_takuma,jpc_takuma,takuma_pci,prakash_pci}.

\begin{figure}
    \centering
    \includegraphics[width=0.95\columnwidth]{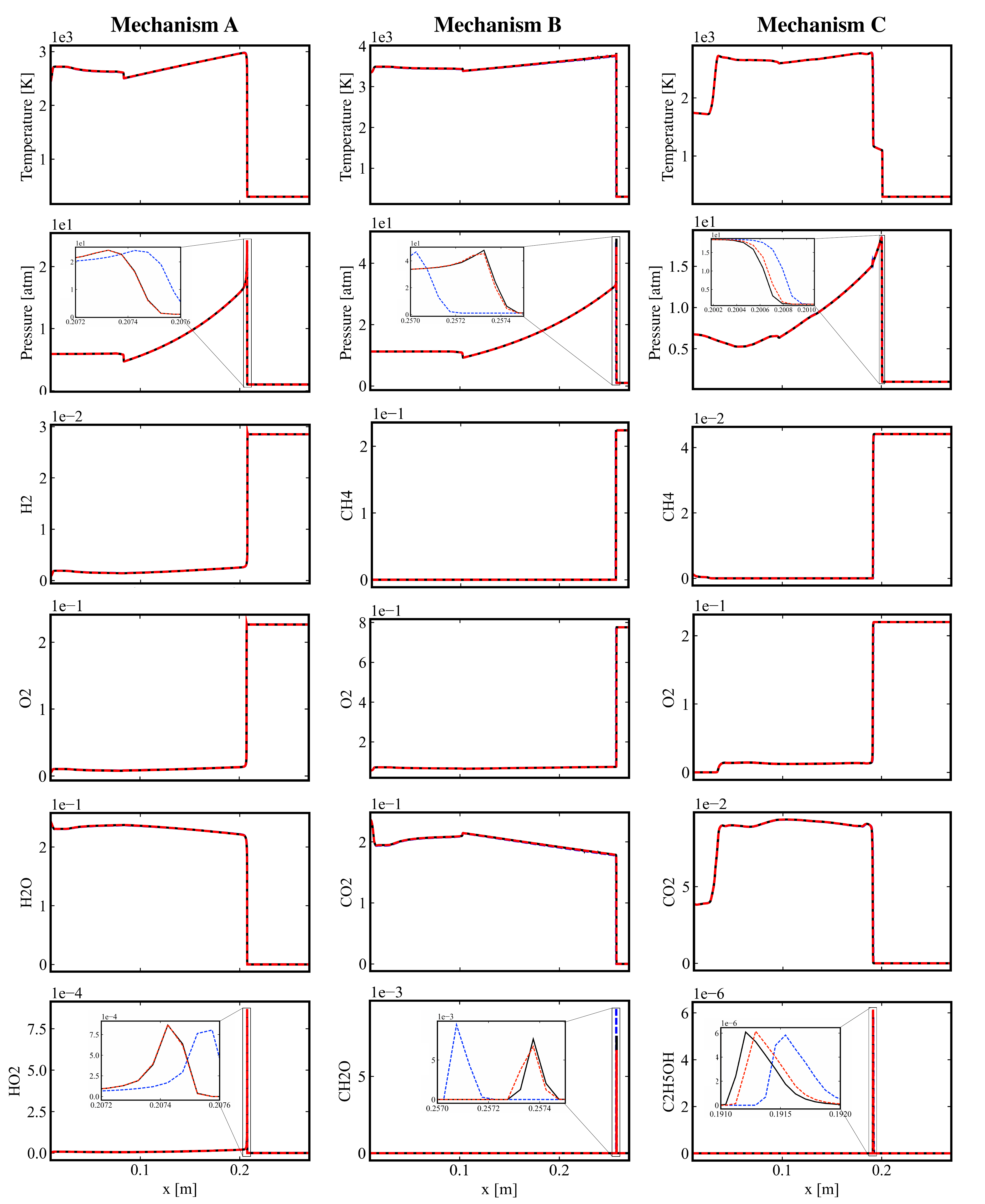}
    \caption{Detonation profiles after $0.1$~ms from exact (\mythickdashedline{red}), LF-ANN (\mythickdashedline{blue}), and HF-ANN (\mythickline{black}) simulations for Mechanism A (left column), B (middle column), and C (right column). First two rows show temperature and pressure profiles; remaining rows show several species mass fraction profiles. Insets show quantities near detonation front.}
    \label{fig:detonation_profiles}
\end{figure}

The pre-detonation (ambient) conditions and other simulation details used for each mechanism case are provided in Tab.~\ref{tab:pre_detonation}. Note that the initial condition for each case contains a small section at the left end of the channel that is filled with a high pressure, high temperature post-detonation mixture to enable detonation propagation. Figure~\ref{fig:detonation_profiles} shows a snapshot of temperature, pressure, and several mass fraction fields collected after 0.1~ms of run time for each mechanism. The results discussed in the 0-D case are in general applicable here: both LF and HF-ANN models almost exactly capture the nonlinear detonation profiles. However, a slight misrepresentation of the the wavefront location obtained by the LF-ANN simulation is observed in the insets in Fig.~\ref{fig:detonation_profiles}, which is a direct consequence of the higher training errors discussed in Sec.~\ref{sec:training}. More specifically, the location of peak pressure is overestimated for the Mechanism A and C cases, and underestimated for the Mechanism B case. As expected, the insets show that these inaccuracies are largely eliminated by the HF-ANN architecture, especially for small intermediary species in Mechanism B (last row in Fig.~\ref{fig:detonation_profiles}). 

\begin{table}[]
    \centering
    \begin{tabular}{c|c|c|c|c|c|c|c|c|}
                  & \textbf{Fuel} & \textbf{Oxidizer} & $\boldsymbol{\phi}$ & \textbf{T [K]} & \textbf{p [atm]} & \textbf{Length [cm]} & $\boldsymbol{\Delta x}$ \textbf{ [cm]} \\ 
       \textbf{Mech. A} & $\text{H}_2$ & Air & 1.0 & 300 & 1 & 30 & 5e-3 \\
       \textbf{Mech. B} & $\text{CH}_4$ & $\text{O}_2$ & 1.15 & 300 & 1 & 30 & 5e-3 \\
       \textbf{Mech. C} & $\text{H}_2$+$\text{CH}_4$ & Air & 1.0 & 300 & 1 & 50 & 8e-3
    \end{tabular}
    \caption{Pre-detonation (ambient) conditions for each mechanism case. Refer to Tab.~\ref{tab:mechanisms} for mechanism details. Ratio of fuel for Mechanism C is 50:50 $\text{H}_2$:$\text{CH}_4$ by volume. All mechanisms use a simulation time step $\Delta t$ of 1e-10~s, adjusted as needed to satisfy a CFL condition of 0.2.}
    \label{tab:pre_detonation}
\end{table}

\begin{figure}
    \centering
    \includegraphics[width=\columnwidth]{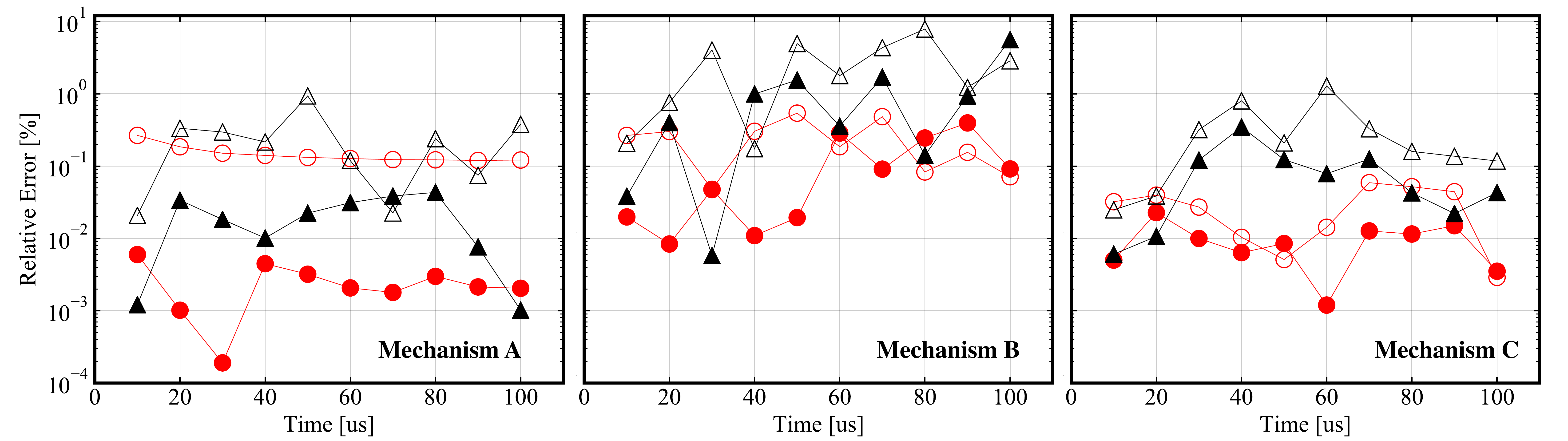}
    \caption{Relative errors in peak temperature (LF-ANN: \mythickbarredcircle{red}{white}, HF-ANN: \mythickbarredcircle{red}{red}) and peak pressure (LF-ANN: \mythickbarredtriangle{black}{white}, HF-ANN: \mythickbarredtriangle{black}{black}) versus time for ANN-based detonation simulations for Mechanisms A (left), B (middle) and C (right).}
    \label{fig:detonation_error}
\end{figure}

Figure~\ref{fig:detonation_error} shows the time evolution of relative errors in peak values for temperature and pressure. Curves for Mechanisms A and C show no constant increase in time, and relative errors for both temperature and pressure stay under 1\% throughout. Further, for most of the snapshots considered in Fig.~\ref{fig:detonation_error}, an expected drop in peak value error is observed when moving from the LF to the HF-ANN model. These trends are also observed for Mechanism B, albeit to a lesser degree: relative errors in peak pressure are as high as 10\% for both HF and LF-ANN models at some time instances, and the amount of overlap between LF and HF-ANN error curves is higher. 


The 1-D detonation results are overall convincing and bring significant confidence towards the viability of the approximate ANN approach. Despite the large increase in complexity over the 0-D cases, the approximate ANNs (both low and high fidelity) capture the relevant detonation structures to acceptable levels of accuracy. However, these results also present a clear trade-off in ANN accuracy versus computational cost, as the increase in ANN fidelity eliminates errors (however small) as expected. Section~\ref{sec:gpu} builds on this demonstration and explores this tradeoff in the context of GPU speedup and saturation.

\subsection{GPU Performance}
\label{sec:gpu}
In this section, the GPU-enabled speedup provided by the formulations in Sec.~\ref{sec:methodology} is explored. In particular, three questions are addressed: 1) For a given $N_C$, how much faster is the GPU-based evaluation of source term than the Cantera-based CPU counterpart? 2) Do the formulations in Sec.~\ref{sec:methodology} allow the GPU to be utilized to its fullest capacity? 3) In which scenario does switching to an approximate ANN for recovering the equilibrium rate constant provide speedup over the exact ANN form (i.e. when is using the ANNs described in Sec.~\ref{sec:approx_ann} "worth it")? It should be stated that below, GPU speedup and performance is assessed only from the perspective of the source term computation in isolation, and not for an entire solver, which is deemed out of scope. This is because GPU-enabled speedup for an entire reacting flow solver can drastically vary depending on a) the chemistry time-integration algorithm, b) GPU treatment of convective/diffusive fluxes, c) GPU treatment of boundary conditions and domain decomposition based communication steps, and d) the amount (and implementation of) CPU-GPU data transfers. Since the methodology in Sec.~\ref{sec:methodology} exists independently from these factors, the GPU speedup and performance trends are also treated independently. 

The methodology described in Sec.~\ref{sec:methodology} was implemented in the GPU setting with a combination of the CUDA and cuBLAS C++ APIs. Calculations used in the analysis below were performed on a single ORNL Summit node consisting of IBM Power9 CPUs and Nvidia V100 GPUs. It should be noted that absolute values of speedup will of course depend on both CPU and GPU architectures as well as the user implementation of GPU functions. Despite this, the GPU computation trends discussed below -- especially with regards to saturation limits and approximate ANN architecture -- are valuable in general. Further, in the context of domain decomposition based approaches used in high-fidelity multi-physics solvers, the GPU is often used to accelerate routines assigned to one or more MPI ranks (or hardware threads of MPI ranks) that operate over some number of cells/nodes in the domain. The speedup-related quantities and figures discussed below are therefore shown as functions of cell number ($N_C$ in Sec.~\ref{sec:methodology}), and increases in $N_C$ can be interpreted as the result of mesh refinement. 

\begin{figure}
    \centering
    \includegraphics[width=\columnwidth]{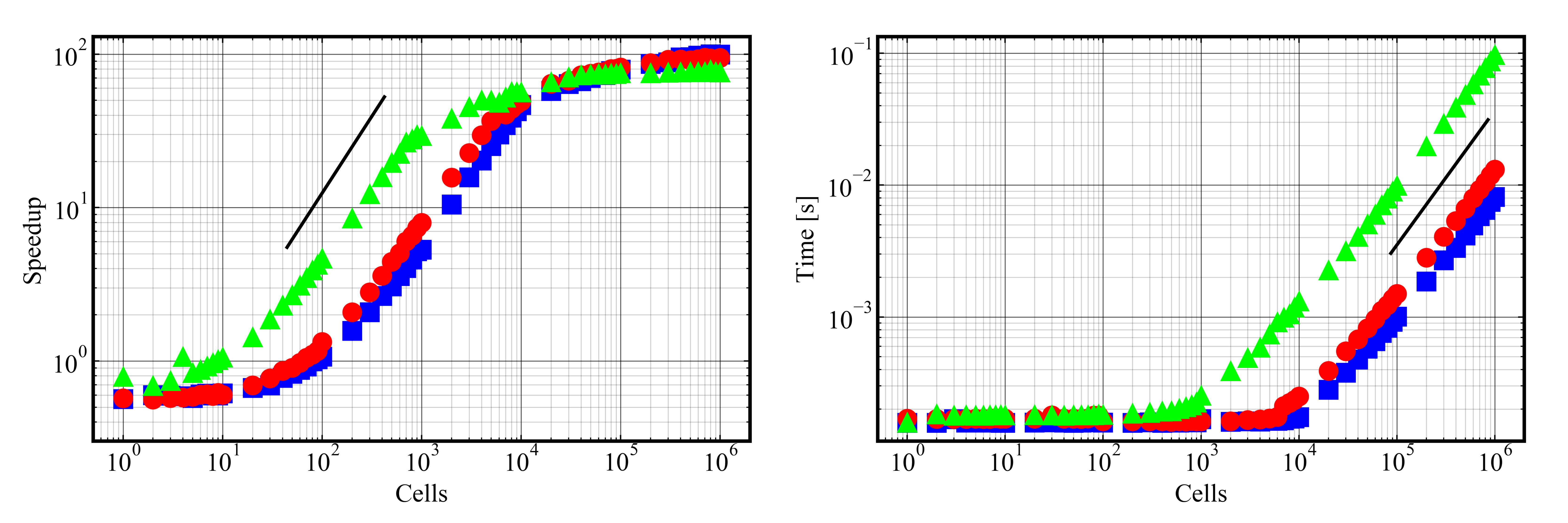}
    \caption{(Left) GPU-enabled speedup in source term calculation over the Cantera-based CPU counterpart versus $N_C$ for Mechanism A (\mysquare{blue}{blue}), B (\mycircle{red}{red}), and C (\mytriangle{green}{green}). (Right) GPU source term evaluation times versus $N_C$ for Mechanism A, B, and C (same symbols). Linear trend with respect to cells (\mythickline{black}) is shown for reference.}
    \label{fig:speedup_exact}
\end{figure}

Figure~\ref{fig:speedup_exact} shows the GPU speedup (ratio between GPU and CPU time-to-solution) and GPU evaluation time for the source term calculation (Eq.~\ref{eq:sourceterm}). The GPU computations use the exact matrix-based formulations from Sec.~\ref{sec:methodology} -- no approximate ANNs are utilized at this point. The CPU baseline used for Fig.~\ref{fig:speedup_exact}a comes from the C++ Cantera function \textit{getNetProductionRates} evaluated with one MPI rank. 

Figure.~\ref{fig:speedup_exact} (left) shows that the speedup curves of all three mechanisms have similar profiles: a convergence in speedup is reached near 100x after an initial period of near-linear growth. There is no decay in speedup after the convergence point is reached. Further, as mechanism complexity increases, speedup also increases within the $10^1$-$10^4$ cell count range which may seem counter-intuitive. On the other hand, the converged speedup at $10^6$ cells drops slightly with increasing $N_S$ and $N_R$. This comes directly from the fact that the saturation point, or the point at which speedup stabilizes, occurs at a lower cell count when mechanisms become more complex (i.e. the increase in $N_S$ and $N_R$ is accounted for by a decrease in $N_C$). This phenomenon is better accessed in the right plot in Fig.~\ref{fig:speedup_exact}, which shows absolute GPU compute times. For a given mechanism, there is a range of cell counts for which compute time does not change at all; the upper bound of this range (which corresponds to the elbow in left plot of Fig.~\ref{fig:speedup_exact}) drops as mechanism complexity increases. Beyond this point, the GPU compute time increases linearly. Since the CPU compute time also increases linearly with respect to $N_C$, no drop in speedup occurs beyond the saturation point as evidenced by Fig.~\ref{fig:speedup_exact} (left).

\begin{figure}
    \centering
    \includegraphics[width=\columnwidth]{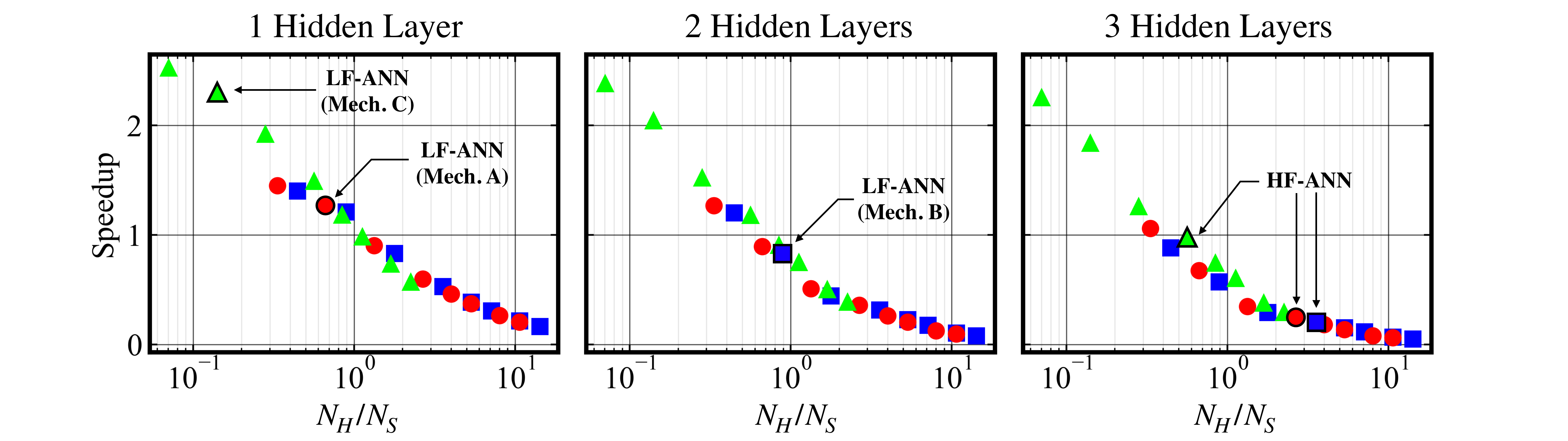}
    \caption{Speedup for GPU-based equilibrium rate constant computation enabled by approximate ANN (Fig.~\ref{fig:trained_ann}, Architecture 2) over the exact form (Fig.~\ref{fig:ann_sourceterm}c) for Mechanism A (\mysquare{blue}{blue}), B (\mycircle{red}{red}), and C (\mytriangle{green}{green}) at $N_C=10^5$. X-axis is number of neurons per hidden layer ($N_H$) normalized by number of species ($N_S$). Plots show increasing number of hidden layers from left-to-right. Arrows indicate LF-ANN and HF-ANN architectures for each mechanism as used in Sec.~\ref{sec:ann_demo}.}
    \label{fig:ann_speedup}
\end{figure}

In Sec.~\ref{sec:ann_demo}, the approximate ANN which replaces the exact computation for the equilibrium constant was utilized. Although the discussion above shows that GPU speedup using the exact matrix-based formulations described in Sec.~\ref{sec:methodology} is high, additional speedup can be extracted through the approximate ANN replacement under some conditions. This is shown in Fig.~\ref{fig:ann_speedup}, where the speedup represents the ratio of times taken to compute $\log(\widetilde{{\bf K}_c})$ (approximate ANN in Fig.~\ref{fig:trained_ann}, Architecture 2) and $\log({\bf K}_c)$ (exact form, Fig.~\ref{fig:ann_sourceterm}c) on the GPU. Note that the speedup shown in Fig.~\ref{fig:ann_speedup} is different than Fig.~\ref{fig:speedup_exact}, which compares GPU-to-CPU ratio in execution times for the exact source terms. 

Figure~\ref{fig:ann_speedup} shows that a) ANN based speedup is affected minimally by the number of hidden layers when the number of neurons per layer is reasonably small, and b) achieving ANN speedup for less complex mechanisms is much less feasible. This second point is especially important, as it signifies how the upper bound on the speedup achievable by the ANN in this context is limited by the number of species and reactions in the mechanism (as evidenced by all mechanisms collapsing to similar curves in Fig.~\ref{fig:ann_speedup}). As a result, even when considering the low-fidelity (LF) ANN models used in Sec.~\ref{sec:ann_demo}, significant speedup is only observed for Mechanism C because the ANN reduction provided is much more significant. Further, despite the fact that the HF-ANNs alleviate much of the errors of the LF-ANNs, none of the HF-ANNs provide speedup over the exact computation of the equilibrium constant. This means the ANN-based speedup also comes at a slight cost in accuracy, though as discussed in Sec.~\ref{sec:ann_demo}, this loss in accuracy does not prohibit the usage of reasonably low-fidelity ANNs in this scope. 

Translation of the approximate ANN based speedup in Fig.~\ref{fig:ann_speedup} into speedup observed in the entire source term calculation (Fig.~\ref{fig:speedup_exact}) is difficult to assess. This is because the values in Fig.~\ref{fig:ann_speedup} (as well as Fig.~\ref{fig:speedup_exact}) for a given ANN architecture are heavily reliant on the GPU implementation of the exact equilibrium constant calculation (Eq.~\ref{eq:log_kc}). This implementation not only affects the ANN-replacement speedup, but also changes the contribution of the equilibrium rate constant computation to the entire GPU-based source term computation. Depending on the implementation, it was found that this contribution can range anywhere from 90\% (inefficient) to 10\% (efficient) of the total source term cost. Despite this variation, in the context of high-fidelity simulations, even a small ANN-derived speedup in Fig.~\ref{fig:ann_speedup} can be useful (i.e. the speedup factor seen for Mechanism A's LF-ANN) as the amount of calls to the source term calculation within one simulation time step is usually very high.

\section{Conclusion}
\label{sec:conclusion}

A method which casts the source term computation into an ANN-inspired form was presented, which allows for an interpretation of the equations as a series of ANN layers. The resulting equations allow for matrix-multiplication based source term estimation where the leading dimension (batch size) can be interpreted as the number of chemically reacting cells in the domain; as such, the approach can be readily adapted in high-fidelity solvers for which an MPI rank offloads the source term computation task to the GPU. Though the exact matrix-multiplication based recasting is GPU-friendly as-is, the ANN-inspired interpretation allows the user to replace portions of the exact routine with trained, approximate ANNs. The ultimate goal is to use these approximate ANNs to decrease computational cost (or increase speedup) over the exact counterparts. 

In this work, the approximate ANNs were trained as drop-in replacements for the equilibrium rate constant computation. The utilization of a trained ANN in this fashion removes the input dependence on the $N_S$-dimensional species concentration vector, which greatly simplifies the training process and allows the framework to maintain certain physical qualities during the source term computation (i.e. Arrhenius form for forward rate constants). Through a-posteriori 0-D auto-ignition and 1-D channel detonation simulations on several mechanisms of varying complexity, the results ultimately showed the viability of this approach in complex nonlinear environments. Despite higher MSE values observed in the training phase, the low-fidelity ANN approximations still produced near-exact mass fraction profiles across the board. 

When the number of cells is reasonably high, the exact GPU-based methodology displays significant speedup over the Cantera CPU counterpart. Further, saturation trends showed that after a near-linear growth in speedup with respect to the number of cells, a saturation point is reached after which the GPU computation time trends linearly with cell count, and speedup stabilizes as a result. As expected, this saturation point was reached for a smaller cell count when the mechanism complexity increased. Further, it was found that speedup obtained by the approximate ANNs for the GPU-based computation of the equilibrium rate constant depended both on the approximate ANN architecture and the chemical mechanism complexity. Achieving approximate ANN speedup for less complex mechanisms was ultimately much less feasible, as the upper bound on the speedup achievable by the ANN in this context is limited by the number of species and reactions. Realistic approximate ANN speedup over the exact form in this sense was achievable only for the more complex chemical mechanism. In general, the analysis showed how the computational advantage provided by the approximate ANNs can be significant, although trained ANNs used to substitute particular algorithms that can already be cast in GPU-friendly forms should not be interpreted as a catch-all technique for speedup.

There are many ways in which this approach can be extended. Since this technique applies only to the source term estimation, coupling of the method with a GPU-optimal stiff time-integration routine is warranted. Further, the design of an approximate ANN architecture better suited for speedup in less complex mechanisms should be explored. Additionally, adapting the ANN-based interpretation to single or half-precision environments can allow for improved GPU utilization, and faster run times as a result. These topics will be explored in future work.

\section*{Acknowledgements}
This research used resources of the Oak Ridge Leadership Computing Facility, which is a DOE Office of Science User Facility supported under Contract DE-AC05-00OR22725. Assistance pertaining to detonation simulation setup from Supraj Prakash is gratefully acknowledged.

\printbibliography

@book{poinsot_book,
  title={Theoretical and numerical combustion},
  author={Poinsot, Thierry and Veynante, Denis},
  year={2005},
  publisher={RT Edwards, Inc.}
}

@article{hon_mueller,
  title={Flow reactor studies and kinetic modeling of the H2/O2 reaction},
  author={Mueller, MA and Kim, TJ and Yetter, RA and Dryer, FL},
  journal={International Journal of Chemical Kinetics},
  volume={31},
  number={2},
  pages={113--125},
  year={1999},
  publisher={Wiley Online Library}
}

@unpublished{wang_methane,
author = "R. Xu and H. Wang",
year={2018},
title={A reduced reaction model of methane combustion},
note = {Personal communication}
}

@misc{smith_methane,
   author = {G.P. Smith and Y. Tao and H. Wang},
   title = {Foundational fuel chemistry model version 1.0 ({FFCM-1})},
   year = {2016},
   howpublished = "\url{http://nanoenergy.stanford.edu/ffcm1}",
}

@misc{ucsd,
  author = {},
  title = {Chemical-Kinetic Mechanisms for Combustion Applications},
  howpublished = "\url{http://web.eng.ucsd.edu/mae/groups/combustion/mechanism.html}",
  year = {2016}
}

@inproceedings{asme_takuma,
  title={Detailed chemical kinetics based simulation of detonation-containing flows},
  author={Sato, Takuma and Voelkel, Stephen and Raman, Venkat},
  year={2018},
  booktitle={ASME Turbo Expo},
  publisher={ASME},
  venue={Oslo, Norway},
  isbn={978-0-7918-5105-0},
  pages={},
  volume={4A}
}

@inproceedings{jpc_takuma,
	Author = {Sato, T. and Voelkel, S. and Raman, V.},
	Title = {Analysis of detonation structures with hydrocarbon fuels for application towards rotating detonation engines},
	year = {2018},
	booktitle = {Joint Propulsion Conference},
	publisher = {AIAA},
	venue={Cincinatti, Ohio},
	pages={},
	volume={}
	}

@article{malik-openfoampaper,
  title={A minimally-dissipative low-Mach number solver for complex reacting flows in OpenFOAM},
  author={M. Hassanaly and H. Koo and C. F. Lietz and S. T. Chong and V. Raman},
  Journal={Computer and Fluids},
  Volume={162},
  Pages={11-25},
  year={2018},
}

@article{OF_rhocentral,
  title={Implementation of semi-discrete, non-staggered central schemes in a colocated, polyhedral, finite volume framework, for high-speed viscous flows},
  author={Greenshields, Christopher J and Weller, Henry G and Gasparini, Luca and Reese, Jason M},
  journal={International J. for Numerical Methods in Fluids},
  volume={63},
  number={1},
  pages={1--21},
  year={2010},
  publisher={Wiley Online Library}
}

@misc{goodwin2012cantera,
   author = {David G. Goodwin and Harry K. Moffat and Raymond L. Speth},
   title = {Cantera: an object-oriented software toolkit for chemical
             kinetics, thermodynamics, and transport processes},
   year = {2017},
   howpublished = "\url{https://www.cantera.org}"
}

@inproceedings{takuma_pci,
  title={Mixing and detonation structure in a rotating detonation engine with an axial air inlet},
  author={Sato, Takuma and Chacon, Fabian and White, Logan and Raman, Venkat and Gamba, Mirko},
  organization={Accepted to Proc. Combust. Inst.},
  year={2020},
  booktitle={}
}

@inproceedings{prakash_pci,
  title={Numerical simulation of a methane-oxygen rotating detonation rocket engine},
  author={Prakash, Supraj and Raman, Venkat and Lietz, Christopher and Hargus Jr., William and Schumaker, Stephen},
  organization={Accepted to Proc. Combust. Inst.},
  year={2020}
}

@article{niemeyer2014,
  title={Accelerating moderately stiff chemical kinetics in reactive-flow simulations using GPUs},
  author={Niemeyer, Kyle E and Sung, Chih-Jen},
  journal={Journal of Computational Physics},
  volume={256},
  pages={854--871},
  year={2014},
  publisher={Elsevier}
}

@article{niemeyer2018,
  title={Using SIMD and SIMT vectorization to evaluate sparse chemical kinetic Jacobian matrices and thermochemical source terms},
  author={Curtis, Nicholas J and Niemeyer, Kyle E and Sung, Chih-Jen},
  journal={Combustion and Flame},
  volume={198},
  pages={186--204},
  year={2018},
  publisher={Elsevier}
}

@article{sankaran_solver,
  title={Direct numerical simulations of reacting flows with detailed chemistry using many-core/GPU acceleration},
  author={P{\'e}rez, Francisco E Hern{\'a}ndez and Mukhadiyev, Nurzhan and Xu, Xiao and Sow, Aliou and Lee, Bok Jik and Sankaran, Ramanan and Im, Hong G},
  journal={Computers and Fluids},
  volume={173},
  pages={73--79},
  year={2018},
  publisher={Elsevier}
}

@article{rigopoulos_gpu,
  title={A methodology for the integration of stiff chemical kinetics on GPUs},
  author={Sewerin, Fabian and Rigopoulos, Stelios},
  journal={Combustion and Flame},
  volume={162},
  number={4},
  pages={1375--1394},
  year={2015},
  publisher={Elsevier}
}

@article{emerging_trends,
  title={Emerging trends in numerical simulations of combustion systems},
  author={Raman, Venkat and Hassanaly, Malik},
  journal={Proceedings of the Combustion Institute},
  volume={37},
  number={2},
  pages={2073--2089},
  year={2019},
  publisher={Elsevier}
}

@article{menon_ann,
  title={Turbulent premixed flame modeling using artificial neural networks based chemical kinetics},
  author={Sen, Baris A and Menon, Suresh},
  journal={Proceedings of the Combustion Institute},
  volume={32},
  number={1},
  pages={1605--1611},
  year={2009},
  publisher={Elsevier}
}

@article{shivam_ftc,
  title={Data-driven classification and modeling of combustion regimes in detonation waves},
  author={Barwey, Shivam and Prakash, Supraj and Hassanaly, Malik and Raman, Venkat},
  journal={Flow, Turbulence and Combustion},
  pages={1--25},
  year={2020},
  publisher={Springer}
}

@article{pope_isat,
author = {   S.B.   Pope },
title = {Computationally efficient implementation of combustion chemistry using in-situ adaptive tabulation},
journal = {Combustion Theory and Modelling},
volume = {1},
number = {1},
pages = {41-63},
year  = {1997},
publisher = {Taylor & Francis}
}

@article{ranade_ann,
  title={An ANN based hybrid chemistry framework for complex fuels},
  author={Ranade, Rishikesh and Alqahtani, Sultan and Farooq, Aamir and Echekki, Tarek},
  journal={Fuel},
  volume={241},
  pages={625--636},
  year={2019},
  publisher={Elsevier}
}

@article{rigopolous_ANN,
  title={Tabulation of combustion chemistry via artificial neural networks (ANNs): Methodology and application to LES-PDF simulation of Sydney flame L},
  author={Franke, Lucas LC and Chatzopoulos, Athanasios K and Rigopoulos, Stelios},
  journal={Combustion and Flame},
  volume={185},
  pages={245--260},
  year={2017},
  publisher={Elsevier}
}

@inproceedings{christo_ann,
  title={An integrated PDF/neural network approach for simulating turbulent reacting systems},
  author={Christo, FC and Masri, AR and Nebot, EM and Pope, SB},
  booktitle={Symposium (International) on Combustion},
  volume={26},
  pages={43--48},
  year={1996},
  organization={Elsevier}
}

@article{isat_ann,
  title={An economical strategy for storage of chemical kinetics: Fitting in situ adaptive tabulation with artificial neural networks},
  author={Chen, J-Y and Blasco, JA and Fueyo, N and Dopazo, C},
  journal={Proceedings of the Combustion Institute},
  volume={28},
  number={1},
  pages={115--121},
  year={2000},
  publisher={Elsevier}
}

@inproceedings{sharma2020deep,
  title={Deep learning for scalable chemical kinetics},
  author={Sharma, Alisha J and Johnson, Ryan F and Kessler, David A and Moses, Adam},
  booktitle={AIAA Scitech 2020 Forum},
  pages={0181},
  year={2020}
}

@inproceedings{neural_ode,
  title={Neural ordinary differential equations},
  author={Chen, Ricky TQ and Rubanova, Yulia and Bettencourt, Jesse and Duvenaud, David K},
  booktitle={Advances in neural information processing systems},
  pages={6571--6583},
  year={2018}
}

@article{blasco_ann,
  title={Modelling the temporal evolution of a reduced combustion chemical system with an artificial neural network},
  author={Blasco, JA and Fueyo, N and Dopazo, C and Ballester, J},
  journal={Combustion and Flame},
  volume={113},
  number={1-2},
  pages={38--52},
  year={1998},
  publisher={Elsevier}
}

@article{blasco_ann2,
  title={A single-step time-integrator of a methane--air chemical system using artificial neural networks},
  author={Blasco, Javier A and Fueyo, Norberto and Larroya, JC and Dopazo, C and Chen, Y-J},
  journal={Computers and Chemical Engineering},
  volume={23},
  number={9},
  pages={1127--1133},
  year={1999},
  publisher={Elsevier}
}

@article{kempf_ann,
  title={Investigation of lengthscales, scalar dissipation, and flame orientation in a piloted diffusion flame by {LES}},
  author={Kempf, A and Flemming, F and Janicka, J},
  journal={Proceedings of the Combustion Institute},
  volume={30},
  number={1},
  pages={557--565},
  year={2005},
  publisher={Elsevier}
}

@article{kundu_ann,
  title={Application of deep artificial neural networks to multi-dimensional flamelet libraries and spray flames},
  author={Owoyele, Opeoluwa and Kundu, Prithwish and Ameen, Muhsin M and Echekki, Tarek and Som, Sibendu},
  journal={International Journal of Engine Research},
  volume={21},
  number={1},
  pages={151--168},
  year={2020},
  publisher={SAGE Publications Sage UK: London, England}
}

@article{nickolls2010gpu,
  title={The GPU computing era},
  author={Nickolls, John and Dally, William J},
  journal={IEEE micro},
  volume={30},
  number={2},
  pages={56--69},
  year={2010},
  publisher={IEEE}
}

@article{niemeyer2014recent,
  title={Recent progress and challenges in exploiting graphics processors in computational fluid dynamics},
  author={Niemeyer, Kyle E and Sung, Chih-Jen},
  journal={The Journal of Supercomputing},
  volume={67},
  number={2},
  pages={528--564},
  year={2014},
  publisher={Springer}
}

@inproceedings{pytorch,
  title={Pytorch: An imperative style, high-performance deep learning library},
  author={Paszke, Adam and Gross, Sam and Massa, Francisco and Lerer, Adam and Bradbury, James and Chanan, Gregory and Killeen, Trevor and Lin, Zeming and Gimelshein, Natalia and Antiga, Luca and others},
  booktitle={Advances in neural information processing systems},
  pages={8026--8037},
  year={2019}
}

@article{elu,
  title={Fast and accurate deep network learning by exponential linear units (ELUs)},
  author={Clevert, Djork-Arn{\'e} and Unterthiner, Thomas and Hochreiter, Sepp},
  journal={arXiv preprint arXiv:1511.07289},
  year={2015}
}

@article{mindthegap,
  title={Mind the gap: Turbulent combustion model validation and future needs},
  author={Hochgreb, Simone},
  journal={Proceedings of the Combustion Institute},
  volume={37},
  number={2},
  pages={2091--2107},
  year={2019},
  publisher={Elsevier}
}

@article{jackie_proc_review,
  title={Petascale direct numerical simulation of turbulent combustion—fundamental insights towards predictive models},
  author={Chen, Jacqueline H},
  journal={Proceedings of the Combustion Institute},
  volume={33},
  number={1},
  pages={99--123},
  year={2011},
  publisher={Elsevier}
}

@article{benedicte_sandia,
  title={Prediction of flame structure and pollutant formation of Sandia flame D using Large Eddy Simulation with direct integration of chemical kinetics},
  author={Jaravel, Thomas and Riber, Eleonore and Cuenot, B{\'e}n{\'e}dicte and Pepiot, Perrine},
  journal={Combustion and Flame},
  volume={188},
  pages={180--198},
  year={2018},
  publisher={Elsevier}
}

@inproceedings{mueller_turnkey,
  title={A computationally efficient turnkey approach to turbulent combustion modeling: From elusive fantasy to impending reality},
  author={Mueller, Michael E},
  booktitle={AIAA Scitech 2019 Forum},
  pages={0994},
  year={2019}
}

@misc{popebook,
  title={Turbulent Flows},
  author={Pope, Stephen B},
  year={2001},
  publisher={IOP Publishing}
}

@article{ramanpitsch-sandia,
  title={A consistent LES/filtered-density function formulation for the simulation of turbulent flames with detailed chemistry},
  author={Raman, Venkatramanan and Pitsch, Heinz},
  journal={Proceedings of the Combustion Institute},
  volume={31},
  number={2},
  pages={1711--1719},
  year={2007},
  publisher={Elsevier}
}

@incollection{menon_lem,
  title={The linear-eddy model},
  author={Menon, Suresh and Kerstein, Alan R},
  booktitle={Turbulent combustion modeling},
  pages={221--247},
  year={2011},
  publisher={Springer}
}

@article{prism,
  title={PRISM: Piecewise reusable implementation of solution mapping. An economical strategy for chemical kinetics},
  author={Tonse, Shaheen R and Moriarty, Nigel W and Brown, Nancy J and Frenklach, Michael},
  journal={Israel Journal of Chemistry},
  volume={39},
  number={1},
  pages={97--106},
  year={1999},
  publisher={Wiley Online Library}
}

\end{document}

